\documentclass[useAMS]{mn2e}
\usepackage{graphicx}
\usepackage{times}
\usepackage{psfig}
\usepackage{epsfig}
\usepackage{longtable}
\usepackage{lscape}
 
\def\kms{$\mathrm{km\;s}^{-1}$} 
 
\def\ha{H$\alpha$}

\def\nii{[N~{\small II}]} 
\def\niip{[N~{\small II}]$\,\lambda6548$} 
\def\niig{[N~{\small II}]$\,\lambda6583$}

\def\FWHM{{\it FWHM\/}}

\def\msun{M$_{\odot}$} 
\def\mbh{$M_{\bullet}$} 
\def\sigmac{$\sigma_c$}
\def\vcirc{$v_c$}
\def\vstar{$v_\star$}
\def\mlstar{$(M/L)_\star$}
\def\mlsun{(M/L)$_\odot$}
\def\lbulge{$L_{\it bulge}$}
\def\hst{{\it HST\/}}

\def\h3{$h_{3}$}
\def\h4{$h_{4}$}

\title[Central black hole of NGC~4435]{NGC 4435: a bulge dominated
galaxy with an unforeseen low mass central black hole\thanks{Based on
observations with the NASA/ESA Hubble Space Telescope obtained at
STScI, which is operated by the Association of Universities for
Research in Astronomy, Incorporated, under NASA contract NAS5-26555.}}

\author[L. Coccato et al.]{
L. Coccato$^{1}$\thanks{E-mail: coccato@astro.rug.nl},
M. Sarzi$^{2}$,
A. Pizzella$^{3}$,
E. M. Corsini$^{3,4}$,  
E. Dalla Bont\`a$^{3}$, and
F.~Bertola$^{3}$\\
$^1$Kapteyn Astronomical Institute, Postbus 800, 9700 AV Groningen, The Netherlands\\
$^2$Physics Department, University of Oxford, Keble Road, 
Oxford, OX1 3RH, UK\\
$^3$Dipartimento di Astronomia, Universit\`a di Padova, 
vicolo dell'Osservatorio 2, 35122 Padova, Italy\\
$^4$Scuola Galileiana di Studi Superiori, via VIII Febbraio 2, 35122 Padova, Italy
}

\begin{document}

\date{Accepted ... Received ...; in original ...}

\pagerange{\pageref{firstpage}--\pageref{lastpage}} 
\pubyear{2005}

\maketitle

\label{firstpage}

\begin{abstract}
We present the ionised gas kinematics of the SB0 galaxy NGC~4435 from
spectra obtained with the Space Telescope Imaging Spectrograph. This
galaxy has been selected on the basis of its ground-based
spectroscopy, for displaying a position velocity diagram consistent
with the presence of a circumnuclear Keplerian disc rotating around a
supermassive black hole (SMBH). We obtained the \ha\ and \niig\
kinematics kinematics in the galaxy nucleus along the major axis and
two parallel offset positions. We built a dynamical model of the
gaseous disc taking into account the whole bidimensional velocity
field and the instrumental set-up. For the mass of the central SMBH we
found an upper limit of $7.5 \cdot 10^6$ \msun\ at $3\sigma$
level. This indicates that the mass of SMBH of NGC~4435 is lower than
the one expected from the
\mbh$-$\sigmac\ ($5\cdot 10^7$ \msun) and near-infrared
\mbh$-$\lbulge\ ($4\cdot 10^7$ \msun) relationships.
\end{abstract}


\section{Introduction}
\label{sec:introduction}

Over the last decade, kinematical studies have proved the presence of
a supermassive black hole (hereafter SMBH) in the centre of about 40
galaxies of different morphological types. For a variety of reasons,
SMBH are suspected to be present in the centres of all galaxies (see
Ferrarese \& Ford 2005 for a review).

The census of SMBHs is now large enough to probe the links between
mass of SMBHs and the global properties of the host galaxies.
The mass of SMBHs correlates with several properties of their hosting
spheroid, such as the luminosity (Kormendy \& Richstone 1995;
Magorrian et al. 1998; Marconi \& Hunt 2003), the central stellar
velocity dispersion (Ferrarese \& Merritt 2000; Gebhardt et al. 2000),
the degree of light concentration (Graham et al. 2001), and the mass
(Haring \& Rix 2004). On the other hand, the SMBH masses do not
correlate with the main properties of discs (Kormendy 2001),
suggesting that formation of SMBHs is linked only to the formation of
the spheroidal component of galaxies. The recent finding of a new
correlation between the central stellar velocity dispersion and the
rotational circular velocity (Ferrarese 2002; Baes et al. 2003;
Pizzella et al. 2005) indicates that the mass of the SMBH could be
also related to the mass of the dark matter halo.

The mass of SMBHs in elliptical and disc galaxies seems to follow the
same scaling relations. However, accurate measurements of SMBH masses
are available for only 11 disc galaxies (Ferrarese \& Ford 2005) and
the addition of new determinations for S0 and spiral galaxies are
highly desirable.
Reliable mass estimates of SMBHs in disc galaxies have been derived
from observations of stellar proper motions in the Milky Way (Ghez et
al. 2003) and spatially resolved kinematics of water vapor masers
(Miyoshi et al. 1995), ionised gas (see Barth 2004 for a review) and
stars (see Kormendy 2004 for a review) in the other galaxies. It is
worth nothing that stellar proper motions can only be measured in our
galaxy, water masers are not common and both the
other two techniques offer merits and limitations.
Stellar dynamical models are powerful as they give information not
only on the mass of the SMBH but also on the orbital structure of the
galaxy. However, both observational and computational requirements are
expensive. The study of the ionised-gas kinematics represents a much
easier way to trace the gravitational potential of galactic nuclei,
since the orbital structure of the gas can be assumed to have a simple
form, namely the one corresponding to a nearly Keplerian velocity
field. However, the gas is more susceptible to non-gravitational
forces and is often found in non-equilibrium configurations.
Therefore the regularity of the ionised gas kinematics has to be
verified observationally (Sarzi et al. 2001; Ho et al. 2002).  The
analysis of position-velocity diagrams (hereafter PVD) of the
emission-line spectra observed with ground-based spectroscopy allows
the identification of those galaxies which are good candidates for
hosting a circumnuclear Keplerian disc (hereafter CNKD) rotating
around a central mass concentration (Bertola et al. 1998). Their PVDs
are characterized by a sharp increase of the velocity gradient toward
small radii and the intensity distribution along the line shows two
symmetric peaks with respect to the centre (Rubin et al. 1997; Sofue
et al. 1998; Funes et al. 2002). These objects are good candidates for
a spectroscopic follow-up with the {\it Hubble Space Telescope\/}
(HST).

In this paper we present and model the ionised-gas kinematics of the
SB0 galaxy NGC~4435 which we measured in spectra obtained with the
Space Telescope Imaging Spectrograph (STIS). This is one of the
galaxies we selected on the basis of ground-based spectroscopy, for
displaying a PVD consistent with the presence of a CNKD rotating
around a SMBH (Bertola et al. 1998). The paper is organized as
follows. The criteria of galaxy selection are presented in
Section \ref{sec:sample}. STIS observations are described and analyzed
in Section \ref{sec:stis}. The resulting ionised-gas kinematics and the
morphology of the dust pattern are discussed in
Section \ref{sec:kinematics}. An upper limit for the SMBH mass of NGC
4435 is derived in Section \ref{sec:n4435}. Finally, results are
discussed in Section \ref{sec:conclusions}.

\section{Sample selection}
\label{sec:sample}

Our sample is constituted by three disc galaxies, namely NGC~2179, NGC
4343 and NGC~4435. We considered them as interesting targets for STIS
because ground-based spectroscopic observations already allowed the
determination of an upper limit for their SMBH mass, \mbh\ (Bertola et
al. 1998).  Moreover, they are characterized by value of \sigmac
$\approx 150$ \kms . This \sigmac\ value is lower than those of most
of the galaxies so far studied with ionised-gas dynamics and higher
than the few galaxies studied by means of water masers (Ferrarese \&
Ford 2005). For this reason, SMBH mass determinations in the proposed
\sigmac\ range would allow a better comparison between data obtained
either with gas or stellar dynamics. 
Finally, our sample galaxies belong to morphological types which are
underrepresented in the sample of galaxies studied so far.

In this paper we present only the results of NGC~4435, since it is the
only galaxy of our sample with smooth and circularly symmetric dust
lanes as well as it is the only sample galaxy with a regular and
symmetric velocity field of the ionised gas. This makes NGC~4435 an
excellent candidate for the dynamical analysis. We defer discussion of
both NGC~2179 and NGC~4343, as well as the estimate of the upper limit
of their SMBH mass to a forthcoming paper (Corsini et al., in
preparation).

NGC~4435 is a large (2.8 arcmin $\times$ 2.0 arcmin, [de Vaucouleurs et
al. 1991, hereafter RC3]) and bright ($B_T=11.74$ [RC3]) early-type
barred galaxy with intermediate inclination ($i=45^\circ$, from RC3
following Guthrie 1992). It is classified SB0$^0$(s) and its total
absolute magnitude is $M_B^0=-19.41$ (RC3), adopting a distance of 16
Mpc (Graham et al. 1999).

\section{STIS observations and data reduction}
\label{sec:stis}

The long-slit spectroscopic observations of NGC~4435 were carried out
with STIS on March 2003 (Prog. Id. 9068, P.I. F. Bertola). STIS mounted the G750M grating centered at
\ha\ with the 0.2 arcsec $\times$ 52 arcsec slit. The detector was the
SITe CCD with 1024 $\times$ 1024 pixels of 21 $\times$ 21
$\mu$m$^2$. No on-chip binning of the detector pixels yielded a
wavelength coverage between about 6290 and 6870 \AA\ with a reciprocal
dispersion of 0.554 \AA\ pixel$^{-1}$. The instrumental resolution was
1.6 \AA\ (FWHM) corresponding to $\sigma_{\rm instr}\simeq30$ \kms\ at
\ha . The spatial scale was 0.05071 arcsec pixel$^{-1}$.

\subsection{Acquisition images}
\label{sec:acquisition}

Four \hst\ orbits were allocated for observing the galaxy. At the
beginning of the first orbit, two images were taken with the F28X20LP
long-pass filter to acquire the nucleus.
The acquisition images have a field of view of $5.4\times5.4$
arcsec$^2$ and a pixel scale of 0.05071 arcsec. The exposure time was
40 s. The long-pass filter is centered at 7230 \AA\ and has a
\FWHM $=2720$ \AA . It roughly covers the $R$ band.

The first image was obtained by adopting for the nucleus the galaxy
coordinates from the RC3 catalog. The image was boxcar-summed over a
check box of $5\times5$ pixels to find the position of the intensity
peak.
The flux-weighted centre of the brightest check box was assumed to be
coincident with the nucleus location. This was used to recenter the
nucleus and to obtain the second image. After determining the nucleus
location, a small move was made to centre the nucleus in the slit.
The acquisition images were bias-subtracted, corrected for hot pixels
and cosmic rays, and flat-fielded using {\small IRAF}\footnote{{\small
IRAF} is distributed by NOAO, which is operated by AURA Inc., under
contract with the National Science Foundation} and the STIS reduction
pipeline maintained by the Space Telescope Science Institute (Brown et
al. 2002). Image alignment and combination were performed using
standard tasks in the {\small STSDAS} package.

\begin{figure*}
\begin{center}
\psfig{file=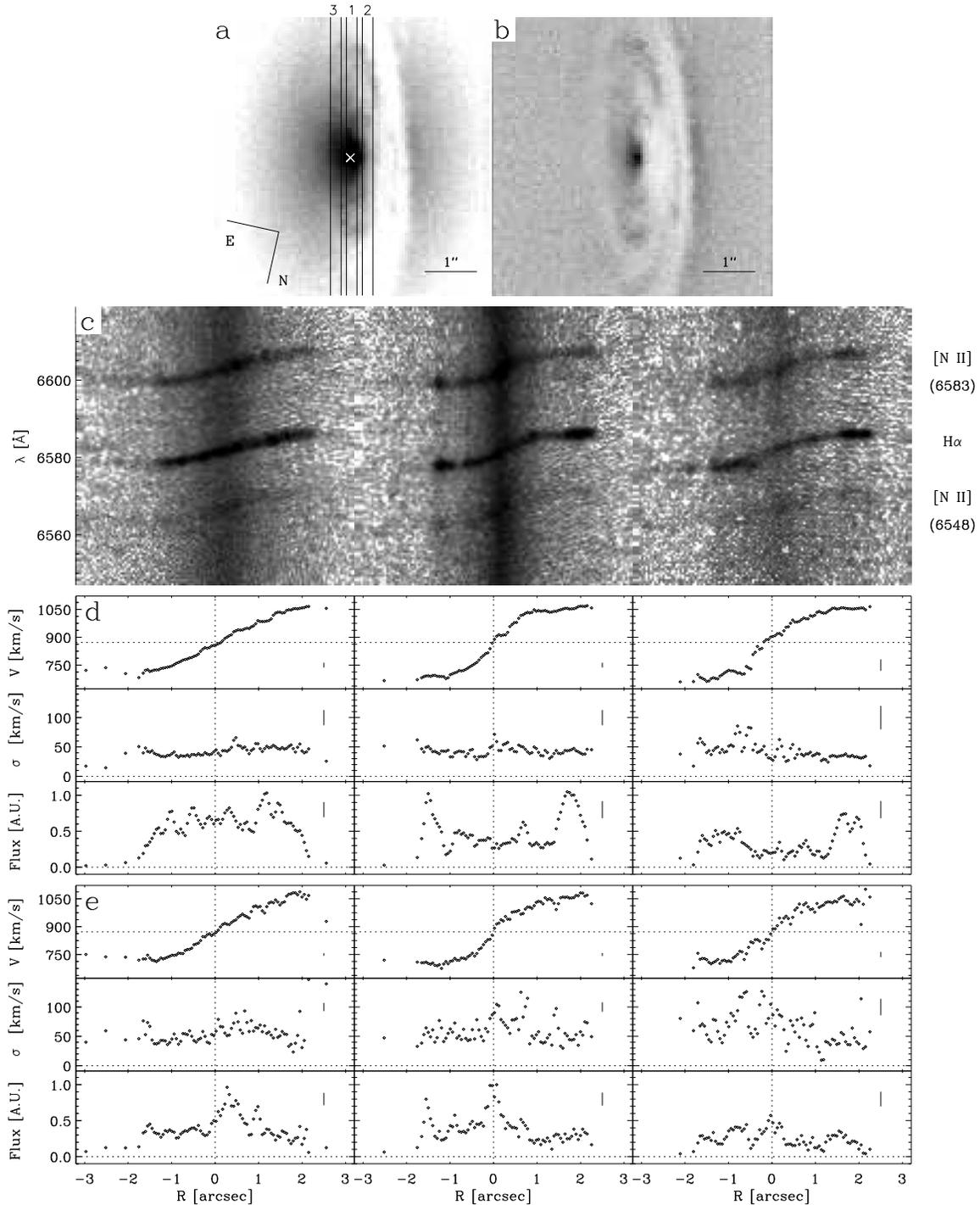,width=16cm,clip=}
\caption{
  {\it a)\/} \hst\ STIS/F28X20LP acquisition image of NGC~4435.  The
  image has been rotated to the STIS instrumental frame. Orientation
  and scale are given. The white cross is the position of the nucleus
  from STIS acquisition procedure. The rectangles overplotted on the
  image show the actual locations of the slit during the spectroscopic
  observations. Table \ref{tab:log} lists the offsets of slit position
  1, 2, and 3 with respect to the location of the nucleus.
  {\it b)\/} Unsharp-masking of the acquisition image of NGC~4435.
  {\it c)\/} Portions of the bidimensional STIS spectra of NGC~4435
  obtained in position 1 ({\it central panel\/}), 2 ({\it right
  panel\/}) and 3 ({\it left panel\/}). The spectral region centered
  on the \ha\ emission line is shown after wavelength calibration,
  flux calibration and geometrical rectification. The spatial axis is
  horizontal and ranges between $-3.2$ and $+3.2$ arcsec, while the
  wavelength axis is vertical and ranges from 6547 to 6619 \AA
  . Individual emission lines are identified using a key, such that
  \nii\ (6548), \ha , and \nii\ (6583) correspond 
  to \niip , \ha, and \niig , respectively.
  {\it d)\/} \ha\ kinematics from the spectra of NGC~4435 obtained in
  position 1 ({\it central panels\/}), 2 ({\it right panels\/}) and 3
  ({\it left panels\/}). For each slit position the line-of-sight
  velocity curve ({\it top panel\/}), the radial profile of the
  line-of-sight velocity dispersion (uncorrected for instrumental
  velocity dispersion, {\it middle panel\/}), and the radial profile
  of line flux in arbitrary units ({\it bottom panel\/}) are
  given. Errorbars are not plotted for clarity, 
  but the typical errors are shown on the right side of each plot.
  {\it e)\/} Same as in {\it d)}, but for the \niig\ emission line.
}
\label{fig:n4435}
\end{center}
\end{figure*}

We used the resulting image to check the actual position of the slit
during the spectroscopic observation as discussed in
Section \ref{sec:location} and shown in Figure \ref{fig:n4435}a.
Moreover, this image was analyzed to map the dust distribution in the
nuclear region of the galaxy. We constructed the unsharp-masked image
using an identical procedure to Pizzella et al. (2002). We divided the
image by itself after convolution by a circular Gaussian of width
$\sigma=6$ pixels, corresponding to $0.30$ arcsec.
This technique enhances any surface-brightness fluctuation and
non-circular structure extending over a spatial region comparable to
the $\sigma$ of the smoothing Gaussian.
The resulting unsharp-masked image of NGC~4435 is given in Figure
\ref{fig:n4435}b.

\begin{figure*}
\begin{center}
\psfig{file=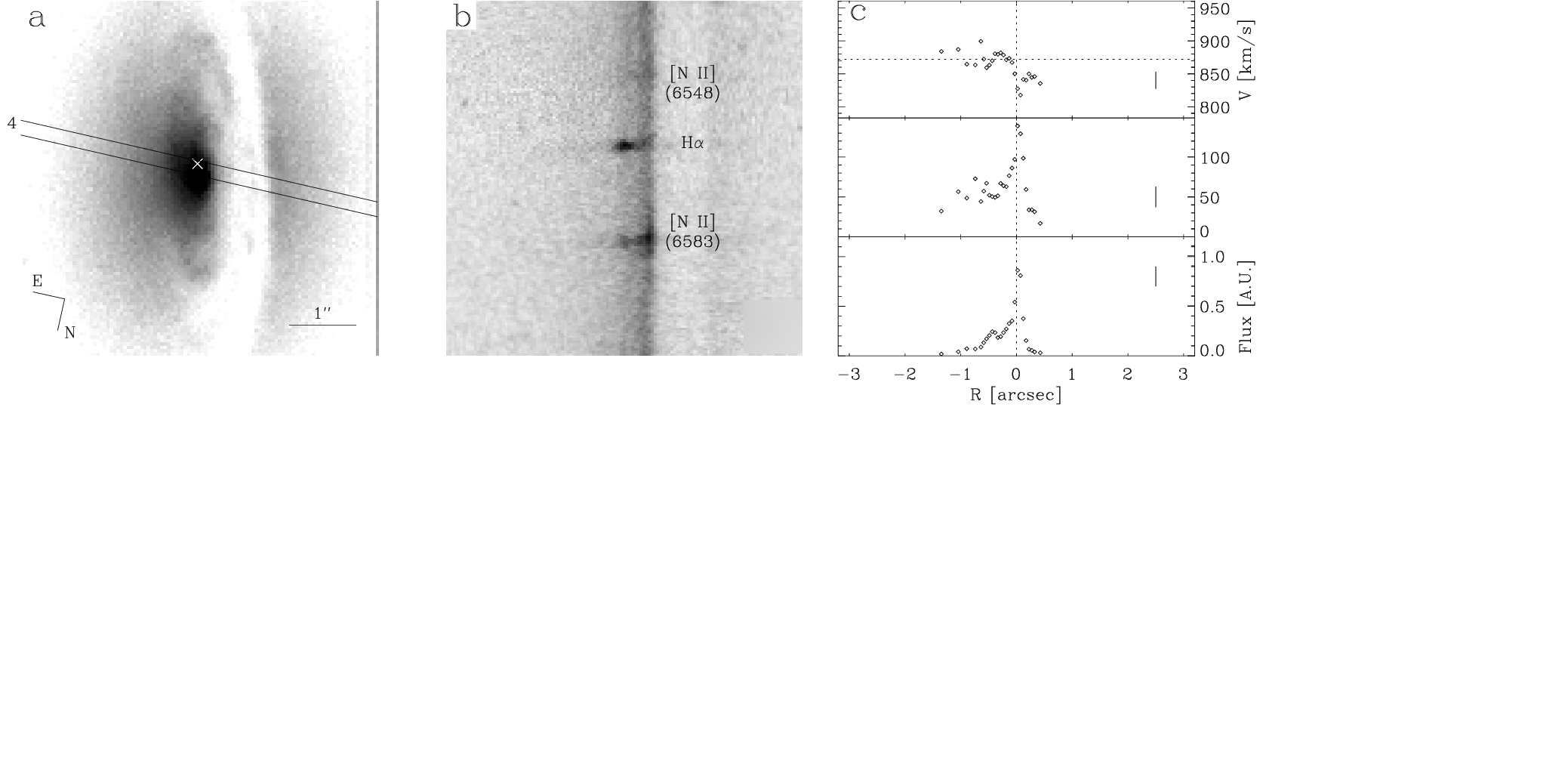,width=13.5cm,bb=53 589 860 858,clip=}
\caption{Similar to Figure \ref{fig:n4435}, 
  but for the archival spectrum of NGC~4435. The acquisition image of
  NGC~4435 with the slit location, a portion of the two-dimensional
  STIS spectrum after data reduction, and the \niig\ kinematics are
  shown in panels {\it a)\/}, {\it b)\/}, and {\it c)\/},
  respectively. In {\it a)\/} the image orientation and scale is the
  same as in Figure \ref{fig:n4435}a. In {\it c)\/} the spatial axis
  ranges between $-3.2$ and $+3.2$ arcsec, and the wavelength axis
  ranges from 6549 to 6627 \AA . }
\label{fig:n4435_min}
\end{center}
\end{figure*}

\subsection{Long-slit spectra}
\label{sec:longslit}

The position angle of the major axis of NGC~4435 is $\rm PA=13^\circ$
(RC3). We took STIS spectra of NGC4435 with the slit centered on the
galaxy nucleus and located along its major axis, and with the slit
parallel to the galaxy major axis on each side of its nucleus with an
offset of $0.25$ arcsec.
Following the target acquisition and peak up, 3 spectra were obtained
with the slit along the major axis. Then the slit was offset westward
by $0.25$ arcsec and 4 spectra were obtained in the first offset
position. Finally, the slit was offset eastward by $0.5$ arcsec and 4
spectra were obtained in the second offset position. Observations of
internal line lamps were obtained during each orbit for wavelength
calibration.
At each slit position, subsequent spectra were shifted along the slit
by 5 detector pixels in order to remove the bad pixels.
The total exposure time for each slit position was balanced within the
constraints of a predefined \hst\ offsetting pattern. The log of the
observations with details about the spectra obtained for NGC~4435 are
given in Table \ref{tab:log}.

The spectra were reduced using the standard STIS reduction
pipeline. The basic reduction steps included overscan subtraction,
bias subtraction, dark subtraction, and flatfield correction.
Subsequent reduction was performed using standard tasks in the {\small
STSDAS} package of {\small IRAF}. Different spectra obtained for the
same slit position were aligned using {\small IMSHIFT} and knowledge
of the adopted shifts along the slit position. Cosmic ray events and
hot pixels were removed using the task {\small LACOS\_SPEC} by van
Dokkum (2001).  Residual bad pixels were corrected by means of a
linear one-dimensional interpolation using the data quality files and
stacking individual spectra with {\small IMCOMBINE}.
We performed wavelength and flux calibration as well as geometrical
correction for two-dimensional distortion following the standard
reduction pipeline and applying the {\small X2D} task. This task
corrected the wavelength scale to the heliocentric frame too.
The contribution of the sky was determined from the edges of the
resulting spectra and then subtracted. The resulting major-axis and
offset spectra of NGC~4435 are plotted in Figure. \ref{fig:n4435}c.

We found in the \hst\ archive 3 other spectra of NGC~4435 which were
taken in 1999 (Prog. Id. 7361, P.I. H.-W. Rix) with the same setup we
adopted for our observations. These spectra were obtained with the
slit placed across the galaxy nucleus with a position angle close to
the galaxy minor axis (see Table \ref{tab:log}). We retrieved the
spectra and reduced them as explained above. The spectrum obtained
close to the minor axis of NGC~4435 is shown in Figure
\ref{fig:n4435_min}c.

\begin{table}
\caption{Log of the STIS observations of NGC~4435. 
  The number of spectra, nominal and actual offset, position angle,
  total exposure time, and observing date are listed for each slit
  position. Slit offsets are given with respect to the location of the
  target peak up.}
\begin{center}
\begin{tabular}{ccrrccr}
\hline
\multicolumn{1}{c}{Pos.} &
\multicolumn{1}{c}{\# Exp.} &
\multicolumn{1}{c}{Nom. Of.} &
\multicolumn{1}{c}{Act. Of.} &
\multicolumn{1}{c}{PA} &
\multicolumn{1}{c}{Exp. T.} &
\multicolumn{1}{c}{Obs. Date} \\
\multicolumn{1}{c}{} &
\multicolumn{1}{c}{} &
\multicolumn{1}{c}{(arcsec)} &
\multicolumn{1}{c}{(arcsec)} &
\multicolumn{1}{c}{($^\circ$)} &
\multicolumn{1}{c}{(s)} &
\multicolumn{1}{c}{} \\
\hline
 1 & 3 &      0  & $+0.05$ &  12.5 & 3012 &  9 Mar 2003 \\
 2 & 4 & $-0.25$ & $-0.25$ &  12.5 & 3574 &  9 Mar 2003 \\
 3 & 4 & $+0.25$ & $+0.30$ &  12.5 & 3584 &  9 Mar 2003 \\
 4 & 3 &      0  & $-0.08$ &  89.6 & 2673 & 26 Apr 1999 \\
\hline
\end{tabular}
\label{tab:log}
\end{center}
\end{table}

\subsection{Location of the slits}
\label{sec:location}

In our observing strategy, the $0.2-$arcsec slit is centered on galaxy
nucleus and it is aligned along the direction of the columns of the
acquisition image at the end of target acquisition and peak up. The
two subsequent offsets were done by applying shifts of $-0.25$
(i.e. westward) and $+0.5$ arcsec (i.e. eastward) in the direction of
the rows of the acquisition image. In the archival spectra slit is
nominally centered on galaxy nucleus.

We determined the actual location of the slits by comparing the light
profile of the spectrum with the light profiles extracted from the
acquisition image.
We obtained the light profile of the spectrum by collapsing the
spectrum along the wavelength direction over the spectral range
between $6350$ and $6800$ \AA .
The comparison profiles were extracted from the acquisition image by
averaging 4 adjacent columns. Each strip corresponds to a synthetic
slit which is $0.2$ arcsec wide.
Each slit position was determined with a $\chi^2$ minimization of the
ratio between the light profile of the spectrum and the light profile
extracted from the acquisition image.

We found that the slit centers were misplaced with respect to their
nominal position. The difference between the nominal and actual
positions of the slit centre with respect to intensity peak of the
acquisition image is typically 1 STIS pixel. The locations of the slit
are overlaid to the acquisition image in Figures.
\ref{fig:n4435}a and \ref{fig:n4435_min}a, and the details about 
their positions are listed in Table \ref{tab:log}.

\subsection{Measurement of the emission lines}
\label{sec:emission}

We derived the kinematics of the ionised-gas component by measuring
the \ha\ and \niig\ emission lines, which are the strongest lines of
the observed spectral range.

For each spectrum we extracted the individual rows out to a distance
of about $2$ arcsec from the slit centre. At larger radii the
intensity of the emission lines dropped off and we therefore binned
adjacent spectral rows until a line signal-to-noise ratio $S/N\geq10$
was attained.
On each single-row extraction we determined the position, FWHM, and
flux of the two emission lines by interactively fitting one Gaussian
to each line plus a straight line to its local continuum.
The non-linear least-squares minimization was done adopting the
{\small CURVEFIT} routine in {\small IDL}\footnote{Interactive Data
Language is distributed by Research System Inc.}.
The centre wavelength of the fitting Gaussian was converted into
heliocentric velocity in the optical convention $v = cz$. The Gaussian
FWHM was converted into the velocity dispersion $\sigma$.
The values of heliocentric velocity and velocity dispersion include no
correction for inclination and instrumental velocity dispersion.
Errors were calculated by taking into account Poisson noise, level of
sky and continuum, read-out noise and gain of the CCD. The noise
associated to the sky level, continuum level, and read out was
estimated from a spectral region free of absorption and emission
features after subtracting the fitted emission lines.
For the archival spectrum of NGC~4435 we fitted only \niig\ line
because the \ha\ line was so weak that little kinematic information
could be derived from it.

Heliocentric velocities, velocity dispersions and fluxes measured from
the \ha\ and \niig\ lines along the major and offset axes of NGC~4435,
are plotted in Figures. \ref{fig:n4435}d and \ref{fig:n4435}e,
respectively. Heliocentric velocities, velocity dispersions and fluxes
measured close to the minor axis of NGC~4435 are given in Figure
\ref{fig:n4435_min}c.

\section{Ionised-gas kinematics and dust morphology}
\label{sec:kinematics}

The presence of a fairly-well defined nuclear disc of dust with
relatively smooth and circularly symmetric dust rings and sharply
defined edges is clearly visible in Figure \ref{fig:n4435}b. It was
first recognized in NGC~4435 by Ho et al. (2002).

We measured the \ha\ and \niig\ kinematics of NGC~4435 out to about 2
arcsec from the centre along the major and offset axes of NGC
4435 (Figures. \ref{fig:n4435}d and \ref{fig:n4435}e).
The \ha\ and \niig\ velocity curves are regular and consistent within
the errors.  However, strong discrepancies are observed along the
central slit for $0\la r \la 0.5$ arcsec, and along the western offset
position for $-1\la r \la -0.5$ arcsec.
We measured consistent values of velocity dispersion from the  
\ha\ and \niig\ lines, in spite of the larger scatter shown by the 
\niig\ data.

We measured only the \niig\ kinematics along the slit position close
to the minor axis of NGC~4435 (Figure \ref{fig:n4435_min}c). The
measurements extend out to $\sim1.5$ arcsec along the eastern side and
to $\sim0.5$ arcsec along the western side, where the dust lanes are
more prominent.
The minor-axis velocity curve and velocity dispersion profiles are
strongly asymmetric. Our velocities are in agreement within the
errorbars with those measured by Ho et al. (2002) by fitting
simultaneously the \ha\ and the two \nii\ emission lines.

 We note that along all slit positions the gas kinematics obtained
from the \ha\ and \niig\ lines are too different to simply combine
them in luminosity-weighted mean values. In particular, the \ha\
rotation is characterised by a shallower velocity gradient than
\nii. Furthermore, most of the Ha flux appear to come from
circumnuclear regions where low \nii/\ha\ ratios suggest star
formation. On the other hand, the \nii\ flux is much more concentrated
toward the centre. Since the \nii\ emission appears to probe better
the nuclear regions and is characterised by a simpler flux
distribution than \ha, we will take into account only the \nii\
kinematics in our model.

NGC~4435 has smooth and circularly symmetric dust lanes as well as a
regular and symmetric velocity field of the ionised gas. On the
contrary, the other two galaxies of our sample are characterized by an
irregular kinematics and an irregular dust lane morphology (Corsini et
al., in preparation). This finding is in agreement with the results of
Ho et al. (2002), that dust lanes can be used as relatively reliable
predictor of the regularity of the gas kinematics in the nuclear
regions of bulges. This makes NGC~4435 an ideal candidate for
dynamical modeling.

\section{Dynamical model}
\label{sec:n4435}

\subsection{Velocity field modeling}
\label{sec:model}

\subsubsection{Basic steps}
\label{sec:basic}

A model of the gas velocity field is generated assuming that the
ionised-gas component is moving onto circular orbits in an
infinitesimally thin disc located in the nucleus of NGC~4435 around
the SMBH. The model is projected onto the sky plane according to the
inclination of the gaseous disc.
Finally the model is ``observed'' simulating as close as possible the
actual setup of STIS spectroscopic observations. The simulated
observation depends on width and location (namely position angle and
offset with respect to the centre) of each slit, STIS PSF and charge
bleeding between adjacent CCD pixels.
The mass of the SMBH is determined by finding the model parameters
which produce the best match to the observed velocity curve.
This modeling technique is similar to that adopted by Barth et
al. (2001) and Marconi et al. (2003) to analyze STIS spectra obtained
along parallel positions across the nucleus of NGC~3245 and NGC~4041,
respectively.

\subsubsection{Model calculation}
\label{sec:calculation}

Let $(r,\phi,z)$ be cylindrical coordinates and consider the gaseous
disc in the $(r,\phi)$ plane with its centre in the origin.
In the case of a spherical mass distribution, the gas circular
velocity \vcirc\ at a given radius $r$ is 
\begin{equation}
v_c(r)= \left[ \frac{GM(r)}{r}\right]^{1/2} = 
  \left[ \left(\frac{M}{L}\right)_\star v^2_\star(r) + 
  \frac{GM_\bullet}{r} \right]^{1/2}
\label{eq:vcirc}
\end{equation}
where $M$ is the total mass enclosed by the circular orbit of radius
$r$, \mlstar\ is the (constant) mass-to-light ratio of the stellar
component (and dark matter halo), and \vstar\ is the circular velocity
of radius $r$ for a stellar component with \mlstar $ = 1$. The radial
profile of \vstar\ is derived from the observed surface-brightness
distribution in Section \ref{sec:stellar}.

The velocity dispersion of the gaseous disc is assumed to be isotropic
with a Gaussian radial profile
\begin{equation}
\sigma(r) = \sigma_0 + \sigma_1 e^{-r^2/2 r_\sigma^2}.
\label{eq:sigma}
\end{equation}
Unfortunately, \ha $+$\nii\ imaging at \hst\ resolution is not
available for NGC~4435. This prevented us to build a map of the
surface-brightness distribution of the ionised gas, as done for
example in Barth et al. (2001). We therefore assume that the flux of
the gaseous disc has an exponential radial profile
\begin{equation}
F(r) = F_0 + F_1 e^{-r/r_F}.
\label{eq:flux}
\end{equation}

We now project the velocity field of the gaseous disc on the sky
plane. Let $(x,y,z)$ be Cartesian coordinates with the origin in the
centre of the gas disc, the $y$-axis aligned along the apparent major
axis of the galaxy, and the $z$-axis along the line of sight directed
toward the observer. The sky plane is confined to the $(x,y)$ plane.
If the gaseous disc has an inclination angle $i$ (with $i=0^\circ$
corresponding to the face-on case), at a given sky point with
coordinates $(x,y)$ the observed gas velocity $v(x,y)$ is
\begin{equation}
v(x,y)=v_c(x,y)\sin{i}\cos{\phi},
\label{eq:vobs}
\end{equation}
where
\begin{eqnarray}
y & = & r \cos{\phi}, \\
r & = & \left[ \frac{x^2}{\cos^2 i}+y^2 \right]^{1/2} .
\end{eqnarray}

We assume that the velocity distribution of the gas at position
$(x,y)$ is a Gaussian with mean $v(x,y)$, dispersion $\sigma(x,y)$,
and area $F(x,y)$.

We now take into account the slit orientation. Let $(\xi,\eta,\zeta)$
be Cartesian coordinates with the origin in the STIS focal plane, the
$\xi-$axis aligned with the direction of the slit width, $\eta$-axis
aligned with the direction of the slit length and the $\zeta$-axis
along the line of sight directed toward the observer and crossing the
centre of the gas disc. The STIS focal plane corresponds to the
$(\xi,\eta)$ plane.
The $(x,y)$ and $(\xi,\eta)$ coordinate systems are related by the
transformation
\begin{eqnarray}
x& = & \xi \cos{\theta} - \eta \sin{\theta} \\ 
y& = & \xi \sin{\theta} + \eta \cos{\theta}  
\label{eq:sky2stis}
\end{eqnarray}
where $\theta$ is the angle between the slit direction and the disc
major axis.

At position $(\xi,\eta)$ the flux contribution due to gas with a
line-of-sight velocity $v$ in the range $[v-\delta v/2,v+\delta v/2]$
is given by
\begin{equation}
F(v|\xi,\eta) = \int_{v-\delta v/2}^{v+\delta v/2} 
  \frac{F(\xi,\eta)}{\sigma(\xi,\eta)\sqrt{2\pi}}
  \exp{\frac{[v'-v(\xi,\eta)]^2}{2\sigma(\xi,\eta)^2}} d v'
\label{eq:model}
\end{equation}
where $\delta v$ is the velocity resolution of the model. For a given
line-of-sight velocity $\hat{v}$, $F(\hat{v})$ is the
``monochromatic'' image of the gas velocity field observed at $\lambda
= \lambda_0 (1+\hat{v}/c)$ for a restframe wavelength $\lambda_0$.

The LOSVD predicted by the model at position $(\xi,\eta)$ on the 
focal plane is
\begin{equation}
S(v|\xi,\eta) = F(v|\xi,\eta) \otimes \mathrm{PSF}(\xi,\eta).
\end{equation}
This takes into account the diffraction of light through the
\hst\ and STIS aperture.

\begin{figure}
\begin{center}
\psfig{file=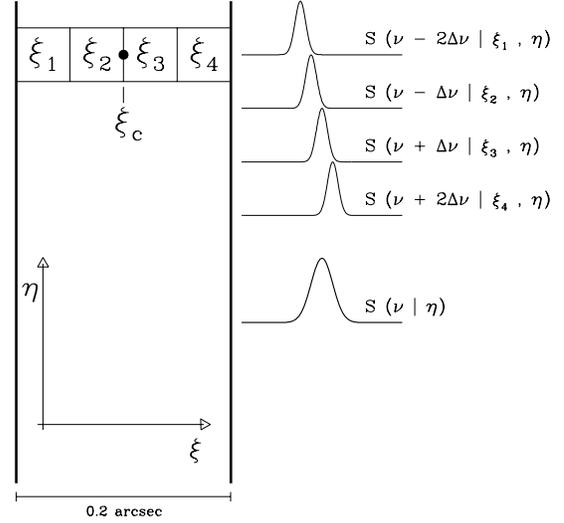, width=8.5cm,clip=}
\caption{Schematic representation of the velocity shift caused by the
  non zero width of the slit. The spectra generated along the slit
  width in pixels of coordinates $(\xi_1,\eta)$, $(\xi_2,\eta)$,
  $(\xi_3,\eta)$, and $(\xi_4,\eta)$ are characterized by a velocity
  offset $\Delta v (\xi_i-\xi_c)$, where $i=1,2,3,4$ and $\xi_c$
  corresponds to the slit centre. The resulting spectrum is
  $S(v|\eta)$.}
\label{fig:shift}
\end{center}
\end{figure}

For each position $\eta$ along the slit the LOSVD predicted by the
model is given by the contribution of all the points on the focal
plane inside the slit
\begin{equation}
S(v|\eta) = \int_{\xi_c-w/2}^{\xi_c+w/2}
S(v+v_d M_a (\xi'-\xi_c) |\xi',\eta) d\xi' 
\end{equation}
where $\xi_c$ is the $\xi$ position of the slit centre, $w$ is the
slit width, $v_d$ is the velocity bin along the wavelength direction
in the spectral range of interest, and $M_a$ is the anamorphic
magnification factor which accounts for the different scale in the
wavelength and spatial direction on the focal plane.
For STIS in the observed spectral range $v_d=25.2$ \kms\
pixel$^{-1}$. The scale in the wavelength direction is $0.05477$
arcsec pixel$^{-1}$ and the scale in the spatial direction is
$0.05071$ arcsec pixel$^{-1}$, thus $M_a=0.93$ (Bowers \& Baum 1998).
The velocity offset $v_d M_a (\xi-\xi_c)$ is the shift due to the
nonzero width of the slit and its projection onto the STIS CCD. The
difference $\xi-\xi_c$ is in pixel units since $v_d$ is given in
\kms\ pixel$^{-1}$. The velocity offset accounts for the fact 
that the wavelength recorded for a photon depends on the position
$\xi-\xi_c$ at which the photon enters the slit along the $\xi$ axis
(Maciejewki \& Binney 2001; Barth et al. 2001). This effect is
sketched in Figure \ref{fig:shift}.

We performed a summation over pixels rather than an integration of
analytic functions.
The model LOSVD and the STIS PSF were calculated on a subsampled pixel
grid with the bin size of $\delta s = 0.01268\times0.01268$ arcsec$^2$
(i.e., a subsampling factor $4\times4$ relative to the STIS pixel
scale) and on a velocity grid with bin size of $\delta v = 10$ \kms .
$\delta s$ and $\delta v$ correspond to spatial and velocity
resolution of the model calculation, respectively. In principle,
smaller values for $\delta s$ and $\delta v$ could give a more
accurate model calculation, but do not increase the result accuracy.
The adopted values are the best comprise between good sampling and
reasonable computational time (300 s on a 1-GHz PC).
We generated the PSF for a monochromatic source at 6600 \AA\ using the
{\small TINY TIM} package in {\small IRAF} (Krist \& Hook
1999). Convolution with the PSF is done using the fast Fourier
transform algorithm (Press et al. 1992).

The model LOSVD was rebinned on a spatial grid with the bin size of
$0.05701\times0.05701$ arcsec$^2$ and on a velocity grid with the bin
size of $v_d=25.2$ \kms\ to match the STIS pixel scale in the spatial
and wavelength direction, respectively.
For each slit position, the array of model LOVSDs forms a synthetic
spectrum which is similar to the STIS spectrum. It was convolved with
the CCD-charge diffusion kernel given by Krist \& Hook (1999) in order
to mimic the bleeding of charges between adjacent STIS CCD pixels.

Finally, we analyzed the synthetic spectrum as the STIS spectra and
measured line-of-sight velocity $v_{\it mod}$, velocity dispersion
$\sigma_{\it mod}$, and flux $F_{\it mod}$ as a function of radius.

\subsection{Stellar component}
\label{sec:stellar}

To investigate the central mass concentration of NGC~4435 it is
necessary to determine the contribution of the stellar component to
the total potential (see Eq. \ref{eq:vcirc}).

\subsubsection{WFPC2 imaging}
\label{sec:wfpc2}

We retrieved Wide Field Planetary Camera 2 (WFPC2) images of NGC~4435
from the \hst\ archive. Data for the filter F450W and F814W
(Prog. Id. 6791, P.I. J. Kenney) were selected to determine the
stellar light profile minimizing the effects of dust absorption.
Two images are available for both filters. Total exposure times were
600 s and 520 s with the F450W and F814W filter, respectively.
All exposures were taken with the telescope guiding in fine lock,
which typically gave an rms tracking error of $0.003$ arcsec.  We
focused our attention on the Planetary Camera chip (PC) where the
nucleus of the galaxy was centered for both the filters. This consists
of $800\times800$ pixels of $0.0455\times0.0455$ arcsec$^2$ each,
yielding a field of view of about $36\times36$ arcsec$^2$.

The images were calibrated using the standard WFPC2 reduction pipeline
maintained by the Space Telescope Science Institute. Reduction steps
include bias subtraction, dark current subtraction, and flat-fielding
are described in detail in Holtzman et al. (1995).
Subsequent reduction was completed using standard tasks in the {\small
STSDAS} package of {\small IRAF}. Bad pixels were corrected by means
of a linear one-dimensional interpolation using the data quality files
and the {\small WFIXUP} task.
Different images of the same filter were aligned and combined using
{\small IMSHIFT} and knowledge of the offset shifts. Cosmic ray events
were removed using the task {\small CRREJ}.
The cosmic-ray removal and bad pixel correction were checked by
inspection of the residual images between the cleaned and combined
image and each of the original frames. Residual cosmic rays and bad
pixels in the PC were corrected by manually editing the combined image
with {\small IMEDIT}.
The sky level ($\sim1$ count pixel$^{-1}$) was determined from regions free
of sources in the Wide Field chips and subtracted from the PC frame
after appropriate scaling.

Flux calibration to Vega magnitudes was performed using the zero
points by Whitmore (1995). To convert to standard $B$ and $I$ filters
in the Johnson system, we estimated the color correction using the
{\small SYNPHOT} package of {\small IRAF} with the S0 spectrum from
Kinney et al. (1996) as template. The color corrections are $B- {\rm
F450W} = +0.115$ and $I- {\rm F814W} = -0.119$.

\subsubsection{Correction of dust absorption}
\label{sec:dust}

We attempted to correct the data for the effects of dust absorption
using a method similar to the one described in Cappellari et
al. (2002). 

\begin{figure}
\psfig{file=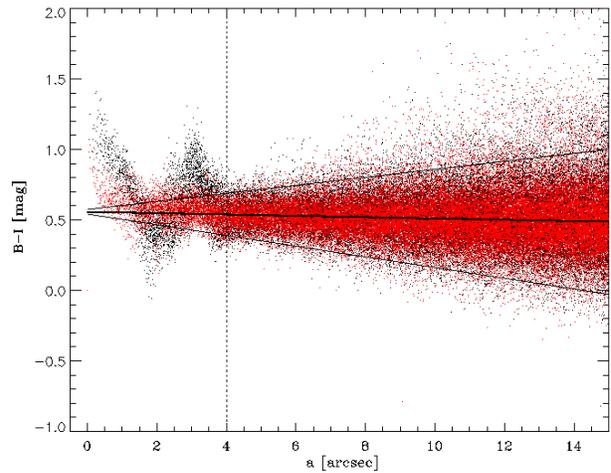,width=8.5cm,clip=}
\caption{Calibrated Johnson $B-I$ color for every pixel
  in the WFPC2 image of NGC~4435 as a function of its elliptical
  radius $a$. {\it Black\/} and {\it red dots\/} refer to pixel on the
  eastern and western side of the galaxy, respectively. The {\it
  thick\/} and two {\it thin continuous lines \/} correspond to fitted
  intrinsic color $(B-I)_0$ and $E(B-I)$ thresholds,
  respectively. Thresholds have been derived for pixels at $4<a<15$
  arcsec and extrapolated to the innermost radii.}
\label{fig:excess}
\end{figure}

\begin{figure*}
\psfig{file=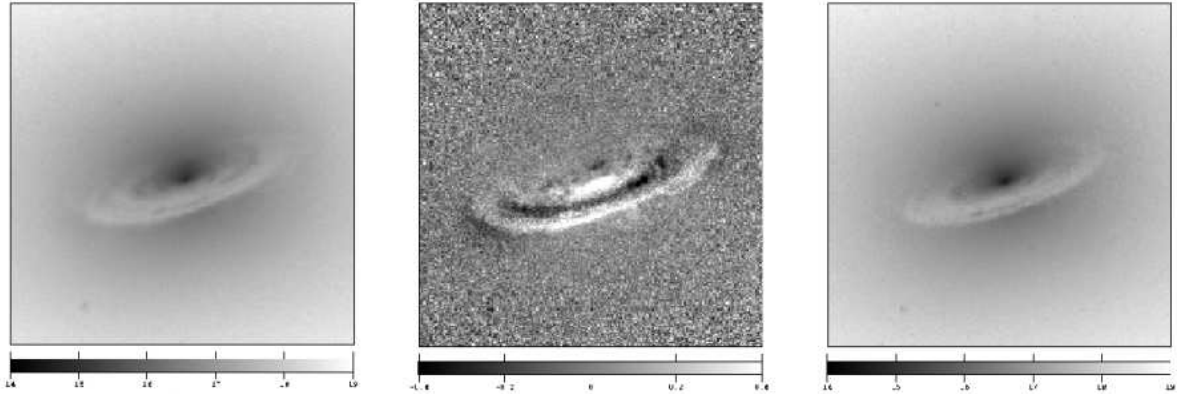,width=16cm,clip=}
\caption{Dust correction of the WFPC2/F814W image of NGC~4435.
{\it Left panel:\/} Observed WFPC2/F814W image after calibration to
Johnson $I-$band. {\it Middle panel:\/} Map of the color excess
$E(B-I)$. {\it Right panel:\/} Extinction-corrected $I-$band image.
The field of view is $10.6\times10.6$ arcsec$^2$ and the
color scale is given in magnitudes.}
\label{fig:dust}
\end{figure*}

For each galaxy pixel we measured the $B-I$ color and we derived the
length $a$ of the semi-major axis of its elliptical isophote.
The average position angle ($\theta = 17^\circ$) and ellipticity
($\epsilon = 0.66$) of the dust features were derived from the analysis
of the unsharp-masked STIS acquisition images as discussed in
Section \ref{sec:orientation}.
We assumed that the intrinsic galaxy color varies linearly as a
function of radius. This assumption is justified by Figure
\ref{fig:excess} which shows the intrinsic galaxy color $(B-I)_0$
obtained as a straight-line fit to the pixel color as a function of
$a$.
For each pixel we computed the color excess $E(B-I)$ as difference
between the measured color $B-I$ and the intrinsic galaxy color
$(B-I)_0$ fitted at that radius. This allowed us to obtain a $E(V-I)$
map of the nuclear region of NGC~4435.  The result is shown in
Figure \ref{fig:dust}. 
We computed the $A(I)=0.558 E(B-I)$ using the standard Galactic
extinction curve given by Cardelli et al. (1989) with the assumption
that the observed color gradient is due to dust rather than stellar
population.
We assumed that dust is distributed in an uniform screen in front of
the galaxy and we used the $A(I)$ map to correct the $I-$band image
for extinction. The extinction-corrected image is shown in Figure
\ref{fig:dust}.
We applied the $A(I)$ correction only to pixels with $|E(B-I)|$ above
a given threshold. By comparing the pixel color in the eastern and
western sides of the galaxy, we defined the threshold as $2$ times the
standard deviation of the observed color with respect to the
intrinsic one. We calculated the threshold for $4<a<15$ arcsec and
extrapolated it for $a\leq4$ arcsec to account for the increasing dust
absorption at smaller radii (Figure \ref{fig:excess}).

This method corrects the major effects of patchy dust absorption.
Nevertheless, the dust disc is still visible in the corrected image,
although it is much less optically thick than in the original one.

\subsubsection{Stellar density profile}
\label{sec:density}

\begin{figure}
\psfig{file=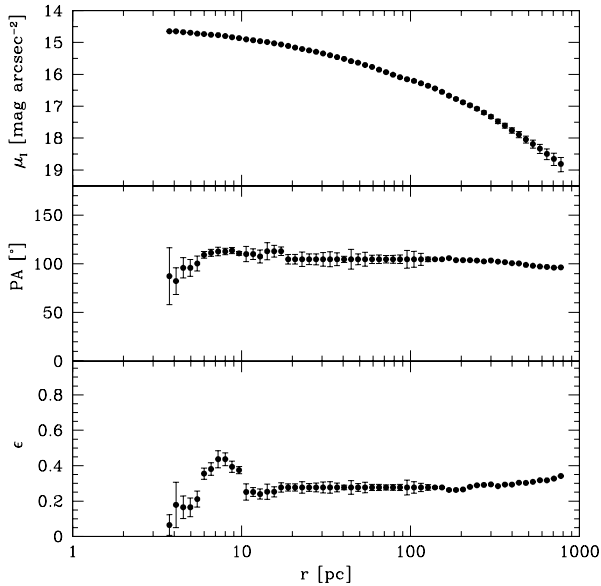,width=8.5cm,clip=}
\caption{Radial profiles of  Johnson $I-$band
 surface brightness ({\it top panel\/}), ellipticity ({\it middle
 panel\/}), and position angle ({\it bottom panel\/}) measured on the
 extinction-corrected image of NGC~4435.}
\label{fig:ellipse}
\end{figure}

Surface photometry was derived on the $I-$band extinction-corrected
image by performing an isophotal analysis with the {\small IRAF} task
{\small ELLIPSE}.
We derived the isophotal profiles of the galaxy by first masking out
the remaining dust patches and then fitting ellipses to the isophotes.
We allowed the centres of the ellipses to vary, to test whether the
light distribution in galaxy nucleus was still affected by dust
obscuration. Since we found some evidence of variations in the fitted
centre, the ellipse fitting was repeated with the ellipse centres
fixed to the location found for the outermost isophotes.  The
resulting azimuthally averaged surface-brightness, ellipticity, and
position angle radial profiles are presented in Figure
\ref{fig:ellipse}.

The isophotes of the masked image are quite circular with
$\epsilon\la0.3$ (Figure \ref{fig:ellipse}). 
For $r<9$ pc (2.5 pixels) the ellipticity is poorly estimated due to
the limited pixel sampling.
This allowed us to treat the surface-brightness distribution as
circularly symmetric, and to assume the stellar density distribution
as spherically symmetric. This approximation is sufficient to estimate
the mass-to-light ratio in the radial range where the ionised-gas
kinematics probes the galaxy potential.

We derived the radial profile of the deprojected stellar luminosity
density $\Gamma(r)$ from the radial profile of the observed
surface-brightness profile $\Sigma(R)$.
The intrinsic surface-brightness profile of the galaxy $S(R)$ and
image PSF were modeled as a sum of Gaussian components using the Multi
Gaussian Expansion (MGE hereafter) described by Monnet et al. (1992)
as done by Sarzi et al. (2001). The PSF for the WFPC2/F814W image was
generated using the {\small TINY TIM} package in {\small IRAF} (Krist
\& Hook 1999).
The multi-Gaussian $S(R)$ was convolved with the multi-Gaussian PSF
and then compared with $\Sigma(R)$ to obtain optimal scaling
coefficients for the Gaussian components. The Gaussian width
coefficients were constrained to be a set of logarithmically spaced
values, thus simplifying the MGE into a general non-negative
least-squares problem for the corresponding Gaussian amplitudes.
For a spherical light distribution, the MGE method leads to a
straightforward deprojection of $S(R)$ into the deprojected stellar
luminosity density $\Gamma(r)$, which can be also expressed as the sum
of set of Gaussians.

\begin{figure}
\psfig{file=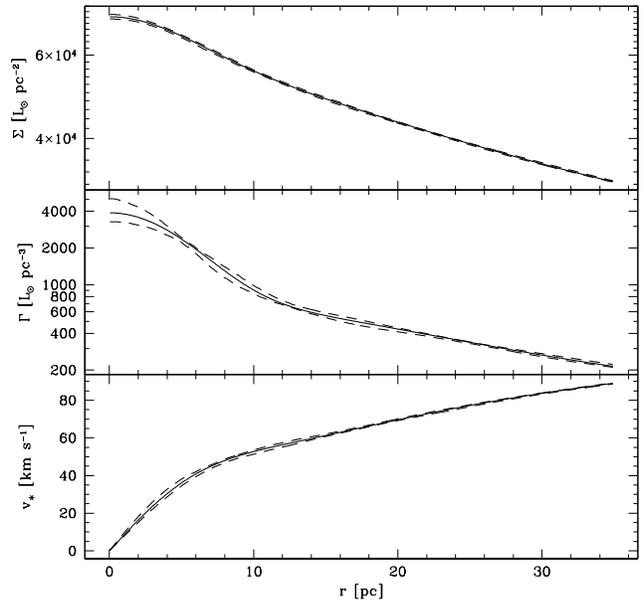,width=8.5cm,clip=}
\caption{Deprojection steps for the stellar mass profile of NGC~4435.
{\it Top panel:\/} Multi-Gaussian fit to observed surface-brightness
profile. {\it Middle panel:\/} Deprojected stellar luminosity density
profile. {\it Bottom panel:\/} Circular velocity curve that results if
the stellar mass and luminosity density are proportional with \mlstar
$ = 1$. In each panel the {\em dashed lines} correspond to the $1\sigma$
confidence limits.}
\label{fig:mge}
\end{figure}

For a spherical mass distribution and a radially constant
mass-to-light ratio, \mlstar , the stellar mass density $\rho(r)$ can
be expressed as $\rho(r)=$\mlstar $\Gamma(r)$. It is the sum of
spherical mass components whose potential can be computed in term of
error functions.
The circular velocity \vstar (r) to be used in Eq. \ref{eq:vcirc} is
derived assuming \mlstar $ = 1$. The multi-Gaussian fit to observed
surface-brightness profile is shown in Figure \ref{fig:mge} along with
the recovered luminosity density profile, and the corresponding
circular velocity curve for \mlstar $ = 1$.

We note that since both B and I-band images are affected by dust,
the B-I maps can only deal a limited description for the dust
distribution. Unfortunately near-IR images of high spatial resolution
for NGC~4435 do not exist.
The use of such images would likely lead to find a larger amount of
dust absorption than presently estimated. 
Hence, we are
currently underestimating the contribution of the stars to the total
mass, and overrating the SMBH mass.
The use of the B-I colour will therefore not affect our conclusions.

\subsection{Orientation of the gaseous disc}
\label{sec:orientation}

The prediction of the observed gas velocity field for NGC~4435 depends
on the orientation of the nuclear gaseous disc, which can be different
from that of the main galaxy disc.

We therefore modeled the gas kinematics by constraining the position
angle $\theta$ and inclination $i$ of the gaseous disc under the
assumption that the dust lanes are a good tracer of the orientation of
the gaseous disc (Sarzi et al. 2001).
We estimated $\theta$ and $i$ by defining ellipses consistent with the
morphology of the dust pattern (as observed in Figure
\ref{fig:n4435}b) and assuming that dust lanes are circularly
symmetric. 
Although the strength of the dust lanes varies spatially, it is
possible to identify the two most conspicuous ones by visual
inspection of the unsharp-masked version of the STIS acquisition
image. To constrain the orientation of the gaseous disc we selected
two of these features. The inner one has a semi-major axis of $\sim2$
arcsec, which corresponds to the maximum extent of our kinematic
measurements, and the outer one marks out the edge of dust pattern on
the western side of the galaxy.

\begin{figure}
\begin{center}
\psfig{file=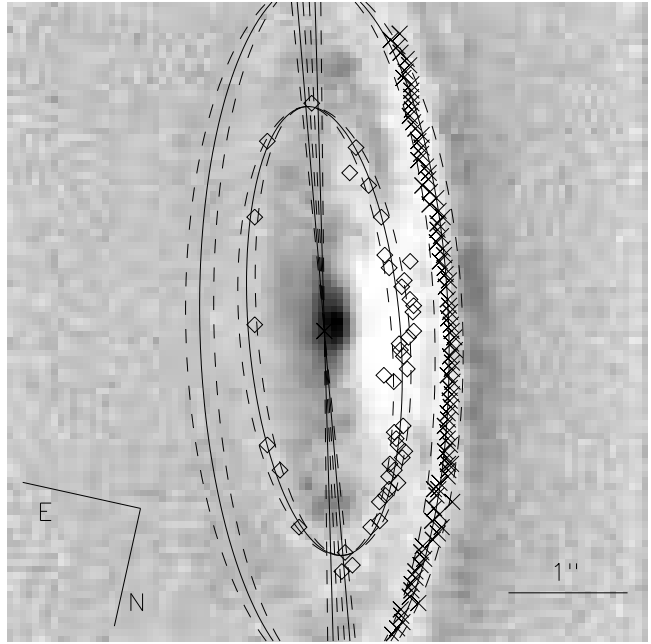,width=8.5cm,clip=}
\caption{The orientation of the gaseous disc of NGC~4435.  {\it
  Diamonds\/} and {\it crosses\/} corresponding to the points of the
  inner and outer dust lanes are overplotted to Figure
  \ref{fig:n4435}b. Absorption features are brighter in this
  unsharp-masked image. The best-fit ellipses of the inner and outer
  dust lanes and their position angles are shown with {\it solid
  ellipses\/} and {\it solid lines\/}, respectively.  The {\it dashed
  ellipses\/} and {\it dashed lines\/} enclose the 1$\sigma$
  confidence ranges on best-fit ellipticity and position angle for
  both the isophotes.}
\label{fig:disc}
\end{center}
\end{figure}

In each row of the unsharp-masked acquisition image we determined the
position of the two main dust lanes by interactively fitting one
Gaussian to each absorption feature plus a straight line to its local
starlight continuum on both side of the nucleus. Once determined the
position of the dust lanes, we fitted them with two ellipses with
same centre (Figure \ref{fig:disc}). The non-linear least-squares
minimization was done adopting the {\small CURVEFIT} routine in
{\small IDL}.
The ellipses have different position angles ($\theta_{\it in} =
17.0^\circ\pm1.0^\circ$, $\theta_{\it out} = 13.8^\circ\pm0.5^\circ$)
but same ellipticity ($\epsilon_{\it in} = \epsilon_{\it out} =
0.66\pm0.03$). The different position angles of the two ellipses can
be interpreted as indicative of a slightly warped gaseous disc, but
this does not affect the final results of the dynamical model as
discussed in Section \ref{sec:bestfit}.
We assumed for the gaseous disc $\theta = 17^\circ\pm1^\circ$ and
$i=70^\circ\pm2^\circ$, since the measured kinematics are encircled by
the ellipse corresponding to the inner dust lane.

\subsection{Best-fitting model}
\label{sec:bestfit}

The parameters in our model are the mass \mbh\ of the SMBH, the
mass-to-light ratio \mlstar\ of the stellar component (and dark matter
halo), the inclination $i$ and position angle $\theta$ of the gaseous
disc, the parameters $\sigma_0$, $\sigma_1$, and $r_\sigma$ of the
Gaussian radial profile of the intrinsic velocity dispersion of the
gas ( see Eq. \ref{eq:sigma}), and the parameters $F_0$,
$F_1$, and $r_F$ of the exponential radial profile of the gas flux
( see Eq. \ref{eq:flux}).
 Given the large number of parameters, it is highly desirable to
constrain as many as possible of them.
We started by fixing the orientation of the gaseous disc using
the results of the analysis of the dust lanes of Section
\ref{sec:orientation}. We therefore assumed $\theta = 17^\circ$ and
$i=70^\circ$ to model the gas velocity field.

We initially considered a grid of models in which every point is
determined by \mbh\ and \mlstar. For every model in this grid we
explored several combination for $\sigma_0$, $\sigma_1$, $r_\sigma$,
$F_0$, $F_1$, and $r_F$ in order to match the observations.
This preliminary analysis revealed that the predicted flux profile
does not depend on the input values of \mbh\ and \mlstar, while the
velocity dispersion profile depends only on the adopted value of
\mbh. The flux parameters need to be adjusted only if different disc
orientations are considered.
This means that we can adopt the same flux parameters ($F_0$, $F_1$,
and $r_F$) for every point of the grid, but we need to adjust the
velocity dispersion parameters ($\sigma_0$, $\sigma_1$, $r_\sigma$)
depending on the value the black hole mass \mbh.
Therefore, we can find the optimal values for $F_0$, $F_1$, and $r_F$
at any position in the $M_\bullet-(M/L)_\star$ grid, by minimizing the
$\chi^2_F = \sum (F - F_{\it mod})^2/\delta F^2$ where $F\pm\delta F$
and $F_{\it mod}$ are the observed and the corresponding model flux
along the major axis (the other slit positions are not
reproducible). The best-fitting values are $F_0=0.08$ and $F_1=0.61$
in the adopted arbitrary units and $r_F=10$ pc.
The velocity dispersion parameters need to be optimized for every
value of the \mbh\ in the grid, by minimizing the $\chi^2_\sigma =
\sum (\sigma -
\sigma_{\it mod})^2/\delta\sigma^2$ where $\sigma\pm\delta\sigma$ and
$\sigma_{\it mod}$ are the observed and the corresponding model
velocity dispersion along the different slit positions, respectively.
The values of the adopted parameters for different \mbh\ values are
listed in Table \ref{tab:sigma_parameters}. Although the velocity
dispersion parameters depend only mildly on the disc orientation, we
changed them when considering different orientations than the one
traced by the dust lanes (Section \ref{sec:disc_orientation}).

\begin{figure}
\begin{center}
\psfig{file=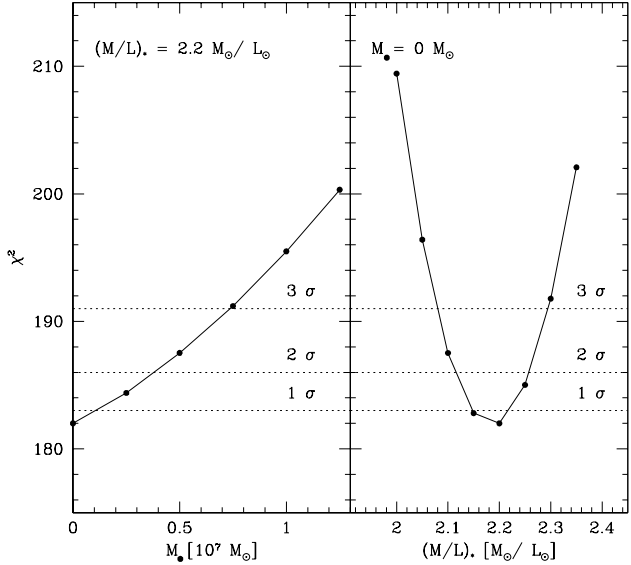,width=8.5cm,bb=20 150 500 590,clip=}
\caption{$\chi^2$ as a function of \mbh\ ({\it left panel\/})
  and \mlstar\ ({\it right panel\/}). The {\it dotted horizontal
  lines\/} indicate the confidence levels on the best fitting values
  of \mbh$ = 0$ \msun\ and \mlstar $=2.2$ \mlsun .}
\label{fig:chisq}
\end{center}
\end{figure}

\begin{figure}
\begin{center}
\psfig{file=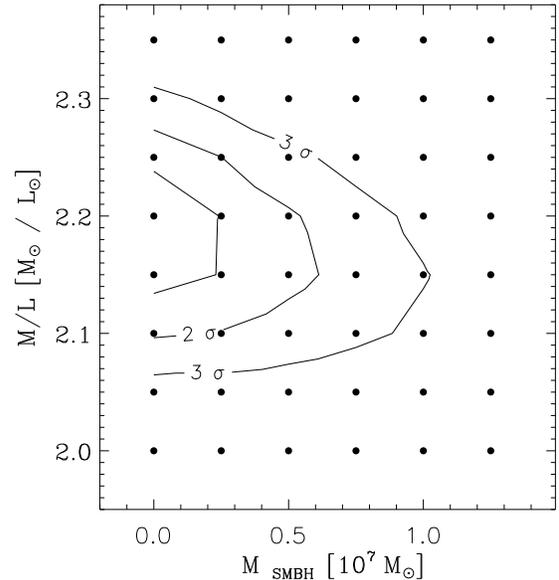,width=8.5cm,clip=}
\caption{$\chi^2$ grid for models with different 
 values of mass-to-light ratio and SMBH mass values.  The {\it continuous
 lines\/} are the confidence levels on the best-fitting values  
 of \mbh$ = 0$ \msun\ and \mlstar $=2.2$ \mlsun .}
\label{fig:grid}
\end{center}
\end{figure}

\begin{figure*}
\begin{center}
\vbox{
 \hbox{
  \psfig{file=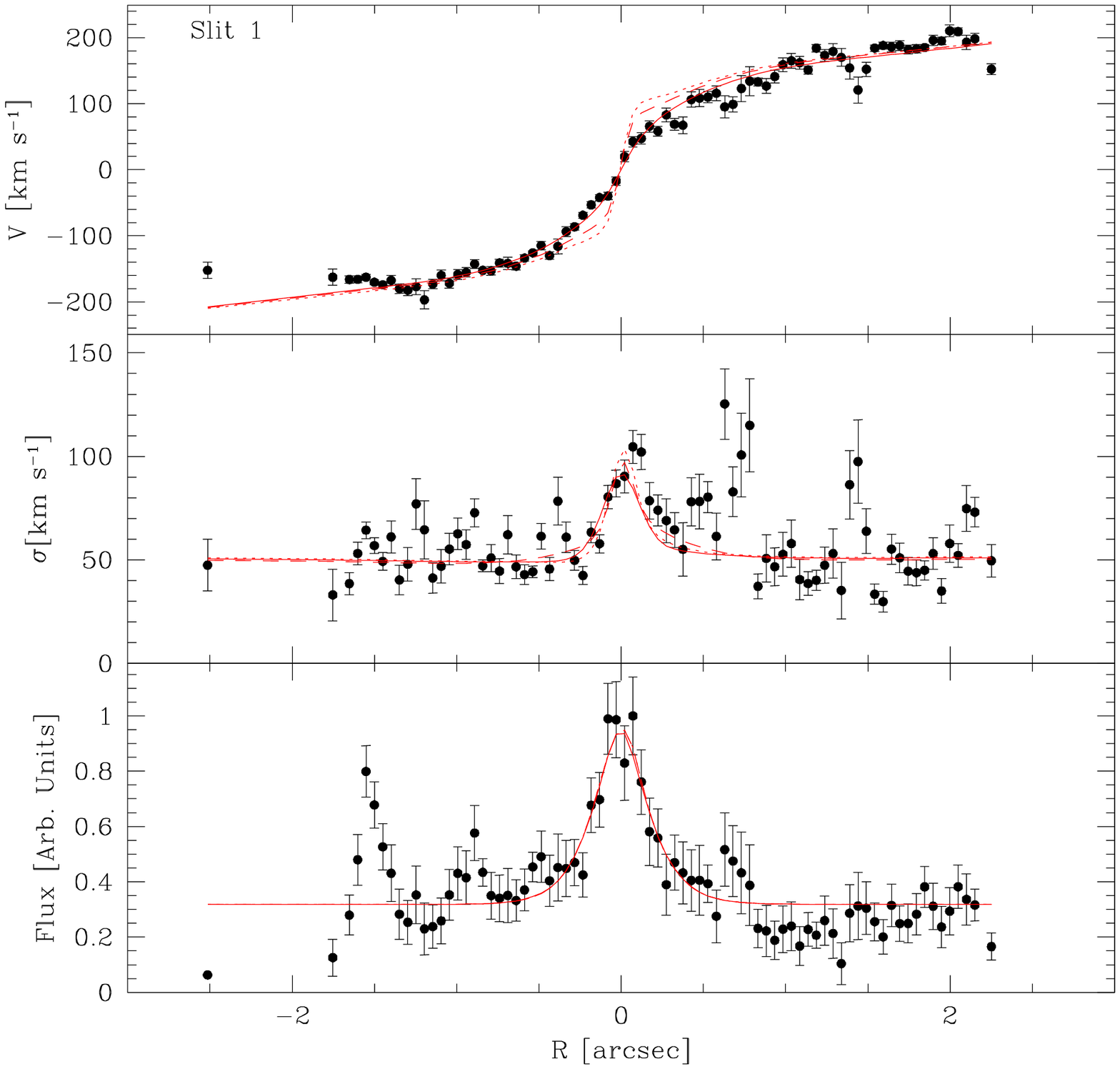,width=8.0cm,clip=}
  \psfig{file=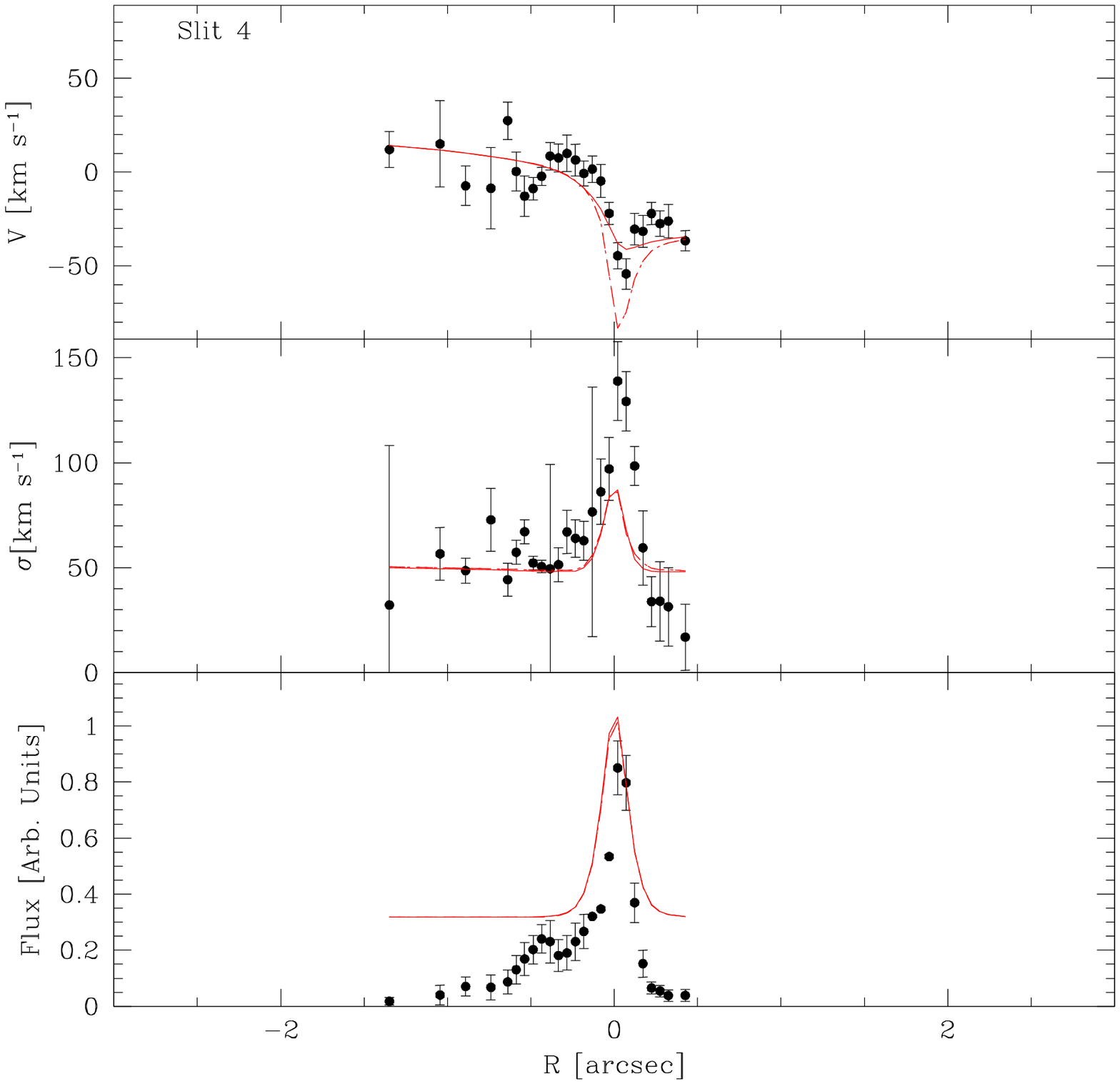,width=8.0cm,clip=}}
 \hbox{
  \psfig{file=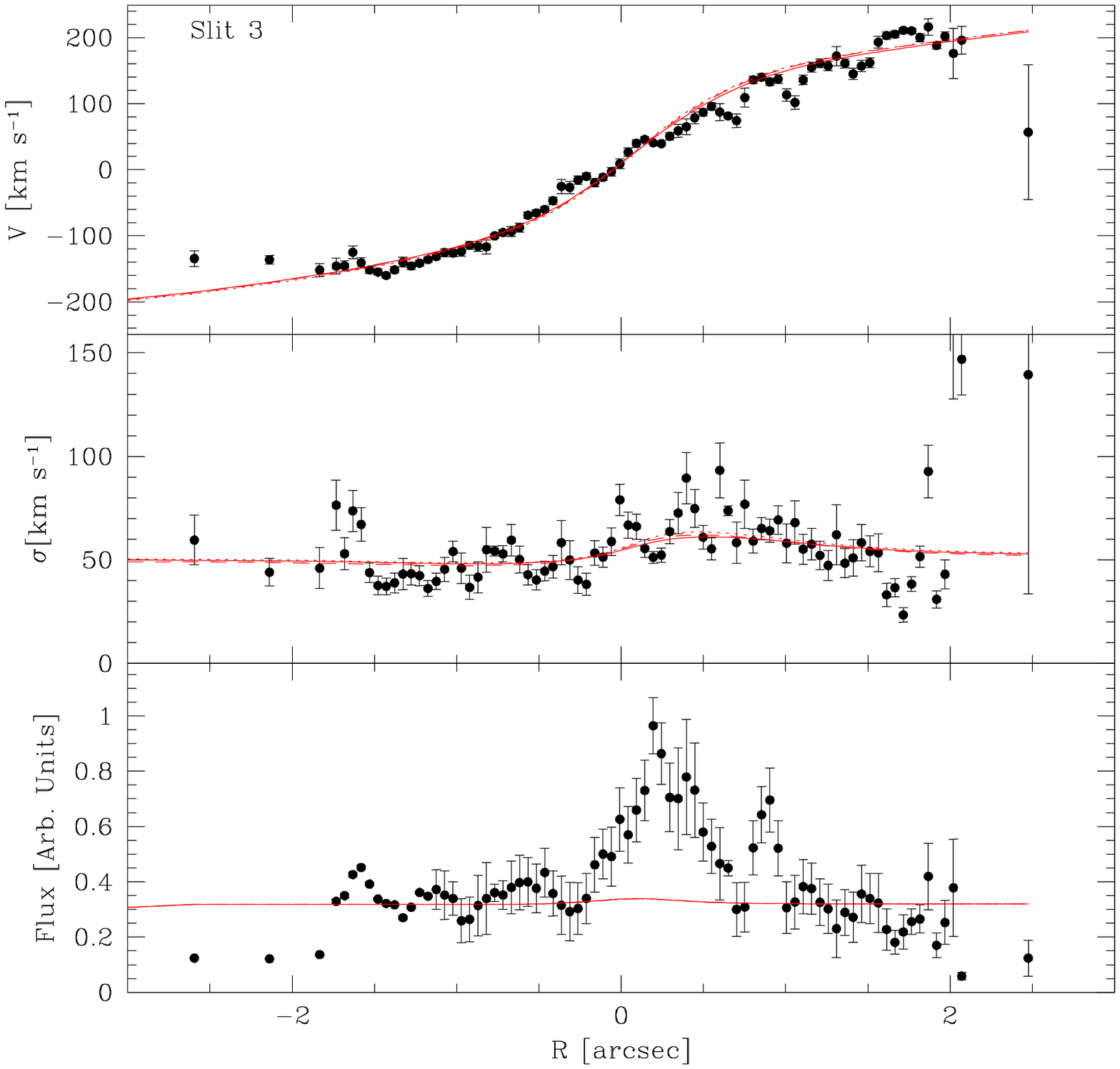,width=8.0cm,clip=}
  \psfig{file=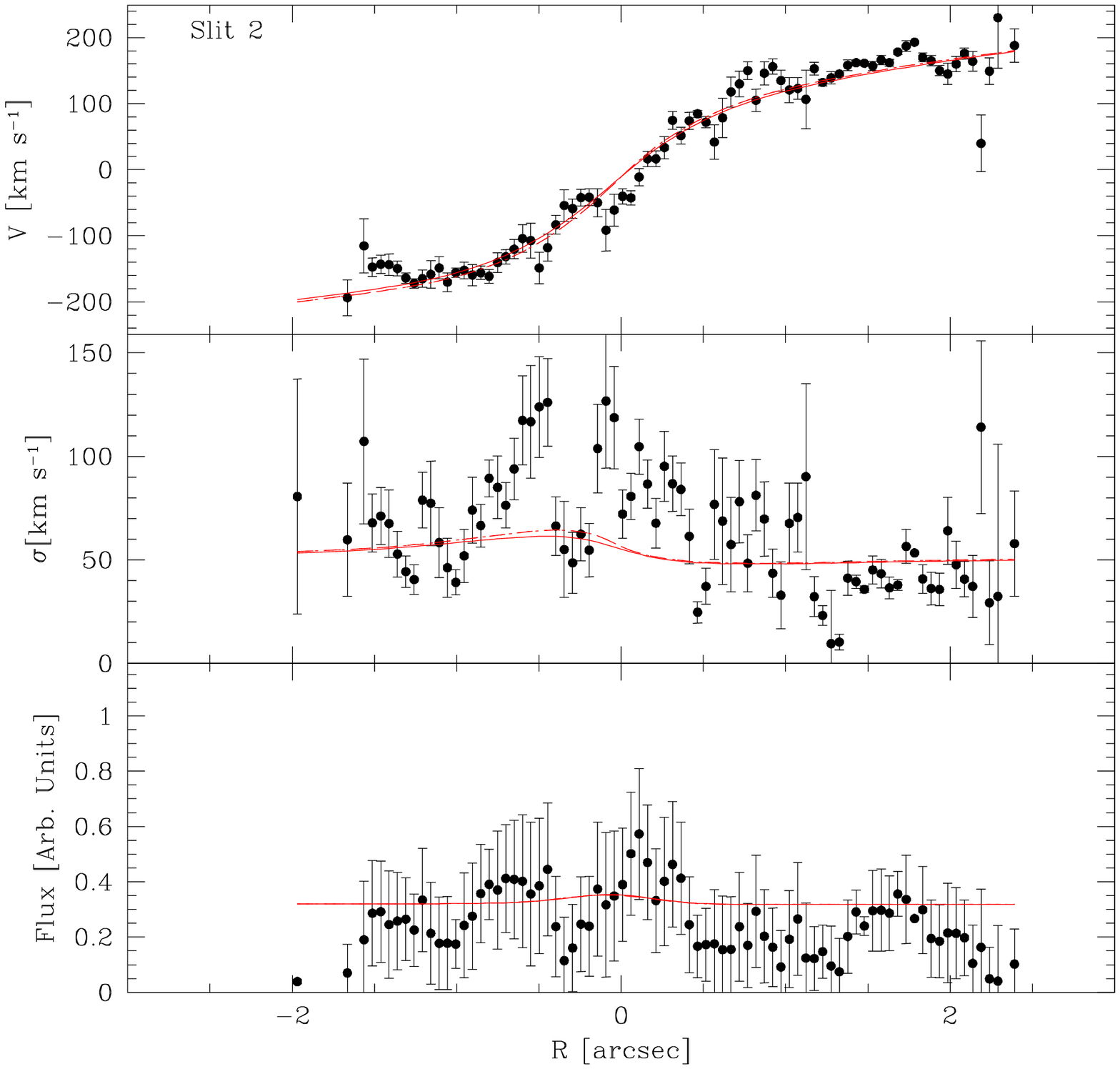,width=8.0cm,clip=}}
}
\end{center}
\caption{Observed \niig\ kinematics ({\it filled circles\/}) along with
  the best-fitting model ({\it solid line\/}) and model predictions
  for the SMBH masses of NGC~4435 derived in Section \ref{sec:bestfit}
  from the \mbh$-$\sigmac\ relation by Ferrarese \& Ford (2005,
  $M_\bullet = 5\cdot10^7$ \msun , {\it dotted line}) and from the
  near-infrared \mbh$-$\lbulge\ relation by Marconi et al. (2003,
  $M_\bullet = 4\cdot10^7$ \msun , {\it dashed line}).  The observed
  and modeled velocity curve ({\it top panel}), velocity dispersion
  radial profile ({\it middle panel}) and flux radial profile ({\it
  bottom panel}) are shown for slit along the major axis (slit \#1,
  {\it upper left panels\/}), close to minor axis (slit \#4, {\it
  upper right panel\/}), along eastern offset (slit \#3, {\it lower
  left panel\/}), and western offset (slit \#2, {\it lower right
  panel\/}).}
\label{fig:fit}
\end{figure*}

We then explored a grid of models with $0 \leq M_\bullet \leq
1.25\cdot10^7$ \msun\ and $2.0 \leq (M/L)_\star \leq 2.35$ \mlsun,
where at every point of the grid we now use the optimized parameters
for the flux and velocity dispersion obtained before. For each model
we calculated the $\chi^2 = \sum (v - v_{\it mod})^2/\delta^2(v)$
where $v\pm\delta(v)$ and $v_{\it mod}$ are the observed and the
corresponding model velocity along the different slit positions.  
The best model has $\chi^2 = 955$ and a reduced $\tilde{\chi}^2 =
5.2$.
The best-fitting model requires no $M_\bullet$ and is compared to the
observed \niig\ kinematics in Figure \ref{fig:fit}.
As we cannot reproduce the observed wiggles and asymmetries, which
could be due to the patchiness of the surface-brightness distribution
of the gas (Barth et al. 2001), in a strict $\chi^2$ sense our best
match is not a good model.
Before deriving confidence limits on \mbh\ and \mlstar, we therefore
need to allow for the velocity structures that are not reproduced by
rescaling all $\chi^2$ values in order for the best $\chi^2$ to match
the number of degree of freedom $N-M$, where $N=184$ is the number of
the observed points and $M=2$ is the number of the parameters in our
model. Note that this procedure is equivalent to allowing for an
additional source of error in our data and leads to more conservative
confidence intervals.
Figure \ref{fig:chisq} show 1$\sigma$, 2$\sigma$, and 3$\sigma$
confidence levels on \mbh\ and \mlstar\ alone, according to the
$\Delta\chi^2$ variations expected for one parameter (i.e. 1, 3, and
9; Press et al. 1992).
Figure \ref{fig:grid} shows similar confidence regions for $M_\bullet$
and \mlstar\ jointly, according to the $\Delta\chi^2$ variation
expected for two parameters (i.e. 2.4, 6.2, 11.8).
For the SMBH in NGC~4435 we derived an upper limit $M_\bullet \leq
7.5\cdot10^6$ \msun\ at 3$\sigma$ confidence level. The mass-to-light
ratio is \mlstar $ = 2.20^{+0.10}_{-0.12}$ at 3$\sigma$ confidence
level.
The sampling of the grid in the parameters spaces is $2.5 \times 10^6
M_\odot$ for the SMBH mass, and 0.05 \mlsun\ for the mass-to-light
ratio. The previous results do not change significantly if we use
a grid with a smaller sampling step.

As stated before, according to the $\chi^2$ analysis the model is
not sophisticated enough to reproduce the data wiggles.  However to
mitigate this problem we can rebin the kinematics in the outermost
regions (beyond $|R|>0.5$ arcsec, where the influence of the SMBH is
negligible). The errors on every re-binned data were computed as the
semi difference between the the maximum value and the minimum value in
the bin. This represent a more reliable estimate of the velocity error
with respect the error derived from the fitting procedure, which take
into account the presence of the wiggles.  We tried several binning on
the observed data in order to obtain the lower value of the reduced
$\chi^2$.  Using this approach we can reach a reduced $\chi^2=1.46$,
with negligible impact on the previous results. The best model still
requires no SMBH and the $3\sigma$ upper limit is now even tighter
($6\cdot10^6$ \msun), a consequence of the augmented relative weight
in the $\tilde{\chi}^2$ of the data close to the center, where the
difference between models with different \mbh\ is the greatest.

\begin{table}
\caption{Parameters of the velocity dispersion
  radial profile adopted in models with \mlstar $=2.2$ \mlsun ,
  $\theta=17^\circ$, $i=70^\circ$, $F_0=0.08$ and $F_1=0.61$ in
  arbitrary units, and $r_F=10$ pc.}
\begin{center}
\begin{tabular}{lccc}
\hline
\multicolumn{1}{c}{\mbh} &
\multicolumn{1}{c}{$\sigma_0$} &
\multicolumn{1}{c}{$\sigma_1$} &
\multicolumn{1}{c}{$r_\sigma$} \\
\multicolumn{1}{c}{[$10^7$ \msun]} &
\multicolumn{1}{c}{[\kms]} &
\multicolumn{1}{c}{[\kms]} &
\multicolumn{1}{c}{[pc]} \\
\hline
0.00  & 40  &    109 &  7.8 \\
0.25  & 40  &    104 &  7.9 \\
0.50  & 40  &    103 &  7.8 \\
0.75  & 40  &    96  &  7.9 \\
1.00  & 40  &    99  &  7.7 \\
1.25  & 40  &    89  &  8.0 \\
4.0   & 39  &    13  &  2.8 \\
5.0   & 41  &    12  &  3.9 \\
\hline
\end{tabular}
\label{tab:sigma_parameters}
\end{center}
\end{table}

Finally, to account for the uncertainty in the estimate of the
inclination and position angle of the gaseous disc we calculated the
best-fitting values of \mbh\ and \mlstar\ by building a grid of models
of the gas velocity fields with $16^\circ \leq \theta \leq 18^\circ$
and $68^\circ \leq i \leq 72^\circ$. We fitted for each disc
orientation the velocity dispersion and flux radial profiles measured
along the major axis. The largest mass for SMBH is \mbh$ \leq 8 \cdot
10^6$ \msun\ at 3$\sigma$ confidence level to find the optimized values
for $F_0$, $F_1$, $r_F$, $\sigma_0$, $\sigma_1$, and $r_\sigma$. The
mass-to-light ratio ranges between \mlstar$ = 2.15^{+0.09}_{-0.12}$
and \mlstar$ = 2.25^{+0.14}_{-0.13}$ at 3$\sigma$ confidence
level. These values are consistent within the errors with the
previously found values.

\section{Comparison with the predictions of the near-infrared 
\mbh$-$\lbulge\ and \mbh$-$\sigmac\ relations}
\label{sec:discussion}

In order to compare our SMBH mass determination for NGC~4435 with the
predictions of the near-infrared \mbh$-$\lbulge\ relation by Marconi
\& Hunt (2003), we retrieved the 2MASS $H-$band images from the
NASA/IPAC Infrared Science Archive. The galaxy image was reduced and
flux calibrated with the standard 2MASS extended source processor
{\small GALWORKS} (Jarrett et al. 2000). We chose the $H-$band image
since it is characterized by lower sky noise with respect to the $J-$
and $K-$band ones.

We performed a bidimensional photometric decomposition using a S\'ersic
law for the bulge component and an exponential law for the disc
component and taking into account seeing smearing. We adopted the
decomposition technique developed by Mendez Abreu et al. (2004). 
The best-fitting parameters are $n = 2.39$, $r_e = 11.14$ arcsec,
$\mu_e = 16.10$ mag arcsec$^{-2}$, and $(b/a)_b = 0.64$ for the bulge
and $h = 14.41$ arcsec, $\mu_0 = 18.15$ mag arcsec$^{-2}$, and
$(b/a)_d = 0.60$ for the disc. The typical error on each parameter is
$<20\%$.  The total bulge luminosity is $L_H = 2.8 \cdot 10^{10}$
L$_\odot$, which corresponds to a \mbh$ = 4 \cdot 10^7$ \msun\
following Marconi \& Hunt (2003).
The presence of the bar in NGC~4435 is enhanced in model-subtracted
image (Figure \ref{fig:decomposition}). However, it has to be noticed
that the bar component dominates the surface-brightness distribution
of NGC~4435 at a larger radial scale with respect to the extension of
STIS kinematics. We conclude that the effects of the bar on the
observed kinematics are negligible in the innermost 2 arcsec.

Several authors report different values for the central stellar
velocity dispersion in NGC 4435.

Bernardi et al. (2002) and Tonry \& Davis (1981) both reported a value
of \sigmac $= 174 \pm 16$ \kms, whereas Simien \& Prugniel (1997)
measured a value of \sigmac $= 156 \pm 7$ \kms. The Hypercat database
also lists an unpublished value of $165 \pm 6$ \kms\ by Prugniel \&
Simien.

In order to be conservative, we adopted the value \sigmac$ = 156 \pm
7$ \kms\ from Simien \& Prugniel (1997). Using the correction
proposed by Jorgensen et al. (1995), we derived the velocity
dispersion that would have been measured within a circular aperture of
$1/8 r_e$, where $r_e$ is the bulge effective radius derived from our
photometric decomposition. We found $\sigma_{1/8}= 157$ \kms. According to
the most recent version of the \mbh$-$\sigmac\ relation (Ferrarese \&
Ford 2005), the expected SMBH mass corresponding to this value of
\sigmac is $5 \cdot 10^7$ \msun.
We note that the other published values of \sigmac would lead to even
larger black-hole mass for NGC4435.

\begin{figure}
\begin{center}
\psfig{file=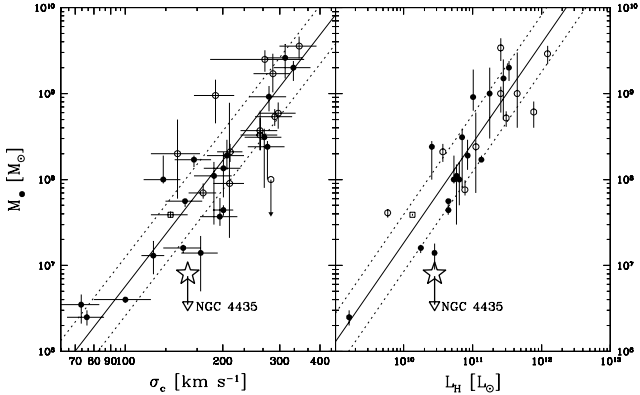,width=8.7cm,bb=25 157 521 470,clip=}
\caption{Location of the upper limit value of the SMBH 
mass of NGC~4435 with respect to the \mbh$-$\sigmac\ relation by
Ferrarese \& Ford (2005, {\it left panel\/}) and near-infrared
\mbh$-$\lbulge\ relation by Marconi \& Hunt (2003, {\it right
panel\/}).
In the left panel, following Ferrarese \& Ford (2005) we plot the SMBH
masses based on resolved dynamical studies of ionised gas ({\it open
circles\/}), water masers ({\it open squares\/}), and stars ({\it
filled circles\/}). The upper limit of the SMBH mass of NGC 4335
(Verdoes Kleijn et al. 2002) has been included for sake of comparison
(see Section \ref{sec:conclusions}).
In the right panel we plot the SMBH masses which for which $H-$band
luminosity of the host spheroid is available in Marconi \& Hunt
(2003).
In both panels {\it dotted lines} represent the $1\sigma$ scatter in
\mbh .}
\label{fig:scaling_relations}
\end{center}
\end{figure}

\begin{figure*}
\begin{center}
\psfig{file=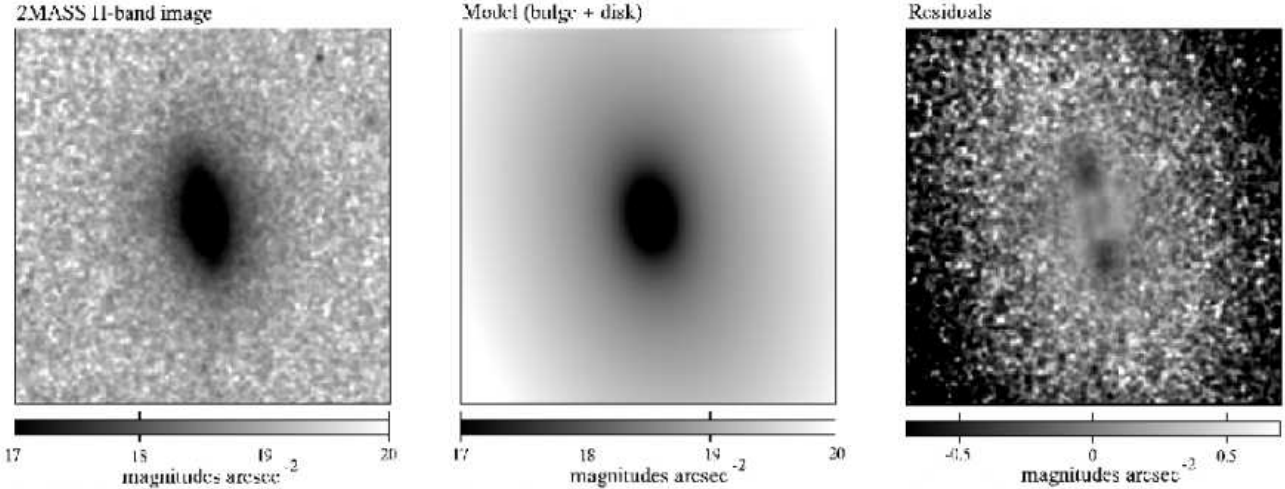,width=18cm,clip=}
\caption{Two-dimensional photometric decomposition of the
surface-brightness distribution of NGC~4435.  {\it Left panel:\/}
Background- and star-subtracted $H-$band image of NGC~4435 from the
2MASS archive. {\it Central panel:\/} Galaxy model as derived from the
photometric decomposition. The bulge is modeled with a S\'ersic law
($n=2.39$, $R_e=11\farcs14$, $\mu_e = 16.10$ mag arcsec$^{-1}$ and
$b/a=0.64$) the disc is modeled with and exponential law
($h=104\farcs41$, $\mu_0=18.15$ mag arcsec$^{-1}$ and
$b/a=0.60$). {\it Right panel:} Residuals of the model. Images are 3 arcmin $
\times$ 3 arcmin.}
\label{fig:decomposition}
\end{center}
\end{figure*}

In both cases, the upper limit found in our model is significantly
below the prediction of these two scaling relations. In Figure
\ref{fig:fit} we show the comparison between the observed
kinematics, the best-fit model, and models adopting the prediction of
the near-infrared \mbh$-$\lbulge\ (Marconi \& Hunt 2003) and
\mbh$-$\sigmac\ (Ferrarese \& Ford 2004) relationships, where the
intrinsic velocity dispersion profiles were accordingly adjusted. The
bigger discrepancies between the observed and predicted velocity
fields are found in the innermost $\pm0.25$ arcsec. In Figure
\ref{fig:scaling_relations} we report the position of the upper limit
of the \mbh\ of NGC~4435 in the \mbh$-$\sigmac\ and near-infrared
\mbh$-$\lbulge\ relations.

\subsection{Uncertainties due the signal-to-noise ratio}

One important point to investigate is if the $S/N$ ratio of our
spectra is sufficiently high to exclude the presence of a SMBH with
the mass predicted by the scaling relations. If it is not the case,
the central velocity gradient is washed out by the noise and it is not
possible to detect the presence of a SMBH.

\begin{figure}
\begin{center}
\psfig{file=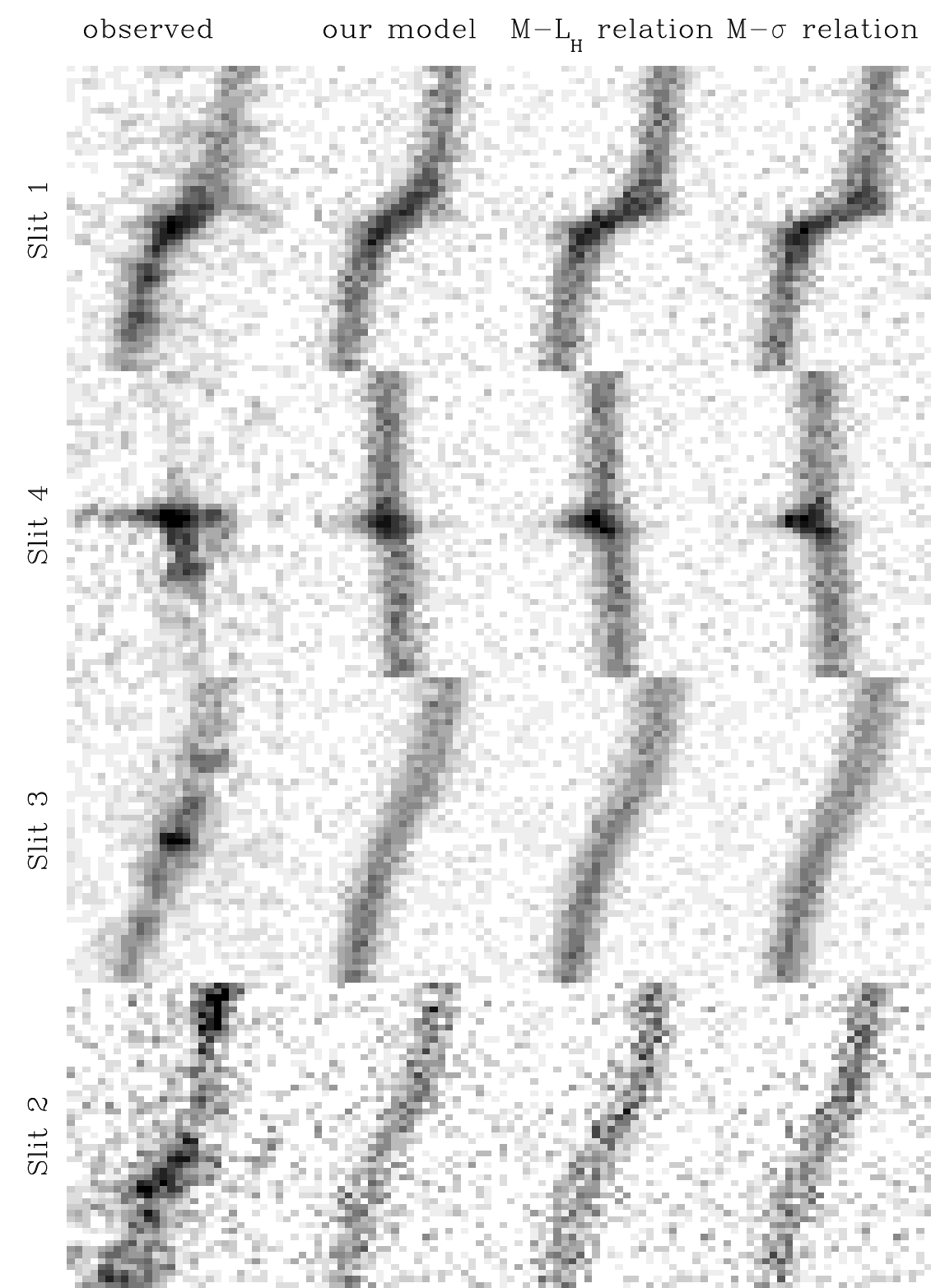,width=7.5cm,clip=}
\caption{Comparison between the observed two-dimensional 
  spectra ({\it left panels\/}) and the two-dimesional synthetic
  spectra for (i) the best-fitting model, (ii) a model with \mbh$ = 4
  \cdot 10^7$ \msun\ as predicted by the near-infrared 
  \mbh$-$\lbulge\ relation, and
  (iii) a model with \mbh$ = 5 \cdot 10^7$ \msun\ as predicted by the
  \mbh$-$\sigmac\ relation for the different slit positions. In the
  models we added a random noise in order to mimic the actual $S/N$
  ratio of the observed spectra. Each box is 15 \AA\ $\times$ 2.6 arcsec.}
\label{fig:spectra}
\end{center}
\end{figure}

To exclude this possibility we measured the noise level in the
observed spectra as the rms of the counts in spectral regions free of
emission lines. This noise has been added to the model spectra in
order to mimic the $S/N$ ratio of the observations.
Figure \ref{fig:spectra} shows that the signature of a SMBH with
\mbh$\geq 4 \cdot 10^7$ \msun\ is clearly visible in the modeled
major-axis spectrum which looks different than the observed one. We
conclude that the velocity gradient corresponding to \mbh$\geq 4
\cdot 10^7$ \msun\ could be measured in the velocity
profiles of the observed spectra.

\subsection{Uncertainties due to the disc orientation}
\label{sec:disc_orientation}

The upper limit we derived for \mbh\ depends on the two assumptions we
did on the orientation of the gaseous disc, i.e.  the morphology of
dust lanes is a good tracer of the orientation of the gaseous disc,
and the gaseous disc is not warped.

\begin{figure}
\centering \psfig{file=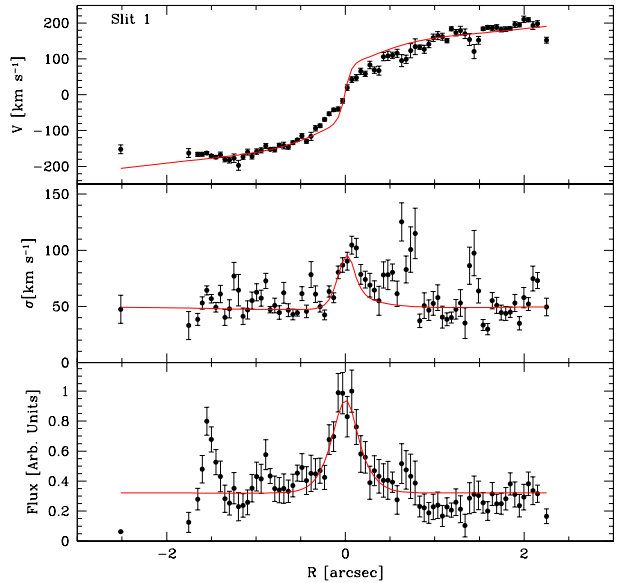,width=8.5cm,clip=}
\caption{Observed \niig\ ({\it filled circles\/}) and fitted 
  ({\it solid line\/}) kinematics in the innermost $\pm2.5$ arcsec
  along the major axis of NGC~4435 obtained with \mbh$ = 4\cdot10^7$
  \msun\ and \mlstar$ = 2.2$ \mlsun\ but allowing the position angle
  and inclination of the gaseous disc to vary. The central velocity
  gradient is not reproduced by any choice of disc inclination and
  position angle.}
\label{fig:patest}
\end{figure}

We test if the presence of a SMBH with a mass of \mbh$ = 4 \cdot 10^7$
\msun\ (i.e., consistent with the near-infrared \mbh$-$\lbulge\
relation) is consistent with the major-axis kinematics assuming for
the gaseous disc a different geometrical configuration with respect to
the one we derived from the dust lanes analysis.  We explored the
space of parameters $\theta$ and $i$ by building a grid of models of
the gas velocity fields with \mbh$ = 4 \cdot 10^7$ \msun\ and \mlstar$
= 2.2$.  For every value of $\theta$ and $i$ we use different
values of $F_0$, $F_1$, $r_F$, $\sigma_0$, $\sigma_1$, and $r_\sigma$,
optimising them according to the prescription given in Section
\ref{sec:bestfit} in order to match the observations.
This value is constrained by the gaseous kinematics measured in outer
regions which is not influenced by the presence of the central
SMBH. The best fit to the observed data is found with $i = 74^{\circ}$
and $\theta = 6.5^{\circ}$. Nevertheless, in the central $\pm0.25$
arcsec there is a great discrepancy between the model and observed
kinematics (Figure \ref{fig:patest}) which convinced us to reject this
model as a reliable solution.

We performed also a simultaneous fit of the velocity curves leaving
the parameters \mbh\, \mlstar, $\theta$ and $i$ free to vary. The best
fit parameters are \mbh $\leq 9 \cdot 10^6$ \msun , \mlstar$ = 2.5 \pm
2$ \mlsun\, $\theta = 15^\circ \pm 2^\circ$, and $i = 75^\circ \pm
3^\circ$.  Although for computational reasons this minimization is
performed with all parameters for the intrinsic flux and velocity
dispersion profiles fixed at the best values derived in Section
\ref{sec:bestfit}, the similarity of the disc orientation found in
this fit with respect to the one determined in Section
\ref{sec:orientation} makes us confident about the choice of adopting
the dust-lane morphology to constrain the orientation of the gaseous
disc.

\begin{figure}
\centering \psfig{file=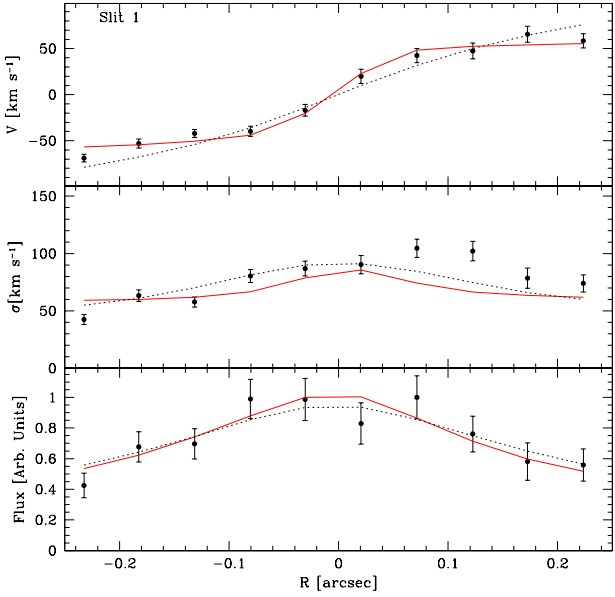,width=8.5cm,bb=18 145 580 690,clip=}
\caption{Observed \niig\ ({\it filled circles\/}) and fitted 
   ({\it solid and dotted lines\/}) kinematics in the innermost
   $\pm2.5$ arcsec along the major axis of NGC~4435. The {\it solid
   lines\/} correspond to a model obtained with \mbh$ = 4 \cdot 10^7$
   \msun , \mlstar$ = 2.2$ \mlsun , $\theta = 10^{\circ}$ and
   $i=28^{\circ}$ in the innermost $\pm2.5$ arcsec. The {\it dashed
   lines\/} correspond to a model obtained with \mbh$ = 0$ \msun ,
   \mlstar$ = 2.2$ \mlsun , $\theta = 17^{\circ}$ and $i=70^{\circ}$.}
\label{fig:best_warp}
\end{figure}

The previous results still rely on the assumption that the gas follows
a coplanar distribution. However, the presence of a SMBH with a mass
\mbh$ = 4 \cdot 10^7$ \msun\ could still be consistent with observed
kinematics if the gas would warp to a more face-on orientation toward
the central region. By exploring the $\theta$ and $i$ parameter space
with models for the gas velocity field only within $\pm0.25$ arcsec
from the centre and with \mbh$=4\cdot10^7$ \msun\ and \mlstar$ = 2.2$
\mlsun, we find that the observed kinematics could be explained if
$i=28^{\circ}$ and $\theta=10^{\circ}$ (Figure \ref{fig:best_warp}).
The images, however, do not suggest such dramatic variation of the
disc orientation, and in Figure \ref{fig:disc} in particular the dust
lanes suggest an highly inclined configuration down to very small
radii ($r\sim0.5$ arcsec).

\subsection{Uncertainties due to the slit position}

The upper limit we derived depends also on the assumed position of the
slits. In Section \ref{sec:location} we determined the actual position
of the slits by comparing the light profile of the spectrum with the
light profiles extracted from the acquisition image. In the case that
this procedure is not reliable, we try to find new slits position (but
not changing their relative distance) in order to allow the presence
of a \mbh$=4\cdot10^7$ \msun. We used the parameters $\theta =
17^{\circ}$ and $i=70^{\circ}$ and \mlstar$ = 2.2$ \mlsun\ and we find
that the major axis velocity gradient is well reproduced if we shift
the slit of 3 {\tt STIS} pixels. But in that case, the kinematics
along the parallel offset slits is not reproduced anymore: there are
no ways to reproduce simultaneously all the slits. Having parallell
slits provide a kinematic test for correct slit positioning, which
confirmed the slit locations inferred in Section \ref{sec:location},
using the acquisition images.

\subsection{Effects of the flux distribution on the measurements}
\label{sec:newFluxDistribution}

Without narrow-band images we had to resort an analytical description
for the intrinsic surface brightness of the ionised gas (see
Section \ref{sec:calculation}). Our choice of an exponential radial
profile for the emission-line fluxes seems well-justified {\em a posteriori}
given the good match to observed flux profiles along all slit
positions, particularly towards the centre (see Figure \ref{fig:fit}).
However, as in the case of NGC~3245 (Barth et al. 2001), it is likely
that unaccounted small-scale flux fluctuations are responsible for the
poor match in many part of the velocity curve (see Section
\ref{sec:bestfit})

As far as the SMBH mass measurement is concerned, it is
important to bear in mind our ignorance of the details characterising
the gas surface brightness towards the very centre of the NGC~4435.
In particular, the presence of nuclear dust may represent a limitation
for our analysis. %
If the very central regions, where the gas clouds are more directly
affected by the gravitational pull of the SMBH, are
affected by significant dust absorption, it possible that gas further
away from the centre and moving at slower speed could contribute to
the observed central velocity gradient and flux peak. If this is
 the case we would be underestimating the SMBH mass.

\begin{figure}
\centering \psfig{file=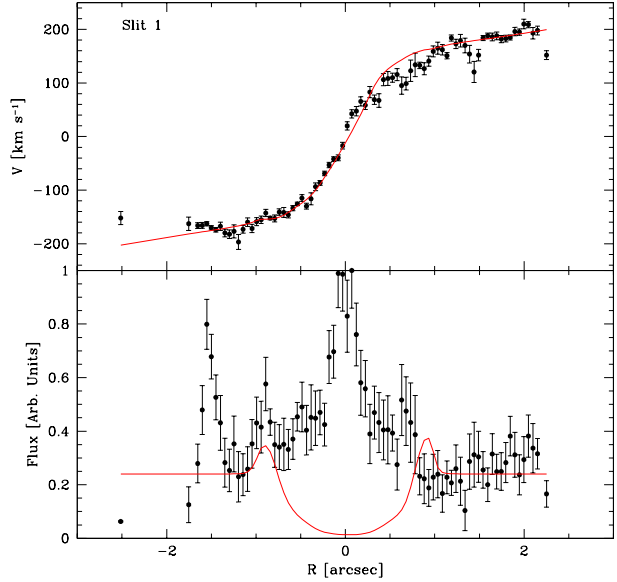,width=8.5cm,clip=}
\caption{Observed \niig\ ({\it filled circles\/}) 
  and fitted ({\it solid lines\/}) kinematics in the innermost
  $\pm2.5$ arcsec along the major axis of NGC~4435. The model is
  obtained using a \mbh=$4\cdot10^7$ \msun\,$i=70^{\circ}$, $\theta =
  17^{\circ}$, \mlstar$ = 2.2$ \mlsun , and an {\it ad hoc} flux
  distribution to reproduce the central gradient. The predicted flux
  radial profile is completely inconsistent with the observed one.}
\label{fig:alternative_flux}
\end{figure}

We investigated this possibility by using a different model for the
intrinsic gas surface-brightness, including an additional central
component with negative amplitude, to mimic a central flux depletion
due to dust absorption. We have started by trying to accomodate a $4
\cdot 10^7$ \msun\ SMBH. As shown by Figure \ref{fig:alternative_flux},
to match the observed gas rotation significant dust absorption is
required out to about 1 arcsec, but in this case the predicted
flux distribution is inconsistent with the observed profile. 
The extent of the nuclear dust is a direct consequence of the fact
that the observed rotation curve can be fitted without a SMBH
(Figure \ref{fig:fit}), so that dust have to screen the emission from
regions where the SMBH can contribute significantly to the observed
rotation curve. For a \mlstar = 2.2, including a $4\cdot 10^7$ \msun\
SMBH still increases the circular velocity (Eq.
\ref{eq:vcirc}) by 10 \kms\ at a radius of 55 pc, corresponding to
$0.7$ arcsec.

In this framework, to explain the observed flux profile we would need
to resort to fine-tuned geometries for the gas distribution in which
intervening off-plane material would contribute to the observed
central peak in flux distribution.
However, the exceptional regularity of the dust distribution in our
images and the well-established connection between dust and gas (e.g.,
Pogge et al. 2000) strongly suggest that the gas resides on a simple a
simple disc.

Furthermore, the required amount of central flux depletion needed to
accomodate a $4 \cdot 10^7$ \msun\ SMBH would translate into an
additional central reddening of $E(B-I)= 1.3$ mag. which is should be
easily detected in our map for the $E(B-I)$ color excess (Figure 5, in
which the central value is $0.6$ mag.). This is not the case,
however. In this calculation we had taken into account i) the
different value of the extinction $A_\lambda$ at different wavelenght
(the $I-$band and the \niig\ region , using the standard Galactic
extinction as done in Cardelli et al. 1989); ii) that only the light
from stars behind the dusty disc is absorbed.

We therefore consider it unlikely that nuclear dust could hide a SMBH
consistent with the expectations of the \mbh$-$\sigmac\ relation.
On the other hand, it is still possible that nuclear dust could impact
our inferred upper limit on the SMBH mass while still
matching the flux profile.  In this case however, the central dust
cannot extend beyond the width of our slit. The impact on \mbh\ will
therefore be rather limited; we find that our upper limit would
increase by only few percent, to $8\cdot 10^6$ \msun. 
Gaussian profile instead of an exponential profile for the flux
distribution, as done by Sarzi et al. (2001, 2002), would have a
similar small effect.

\subsection{Correction for asymmetric drift}

The velocity dispersion of the gas peaks at 100 \kms\ in the centre.
This intrinsic line width is far higher than that expected either from
rotational and instrumental broadening or from thermal motion.
This finding is in agreement with earlier results on bulges of other
disc galaxies based on both ground-based (Fillmore et al. 1986;
Bertola et al. 1995; Cinzano et al. 1999; Pignatelli et al. 2001) and
STIS spectroscopy (Barth et al. 2001) and suggests that random motions
may be crucial for the dynamical support of the gas. If ionised-gas
disc is fragmented into collisionless clouds, then dynamical pressure
supports it against gravity and the mean rotational velocity $v_\phi$
is smaller than the circular velocity $v_c$ given in Equation
\ref{eq:vcirc}. As a consequence, the enclosed mass is underestimated
by dynamical models which do not account for the asymmetric drift
correction.

In the following we address how much of an effect this would have on
the difference between the velocity gradient we measured and that we
inferred for gas rotating around a SMBH with \mbh$ = 4 \cdot 10^7$
\msun\ as predicted by the near-infrared \mbh$-$\lbulge\ scaling relation.

\begin{figure}
\begin{center}
\psfig{file=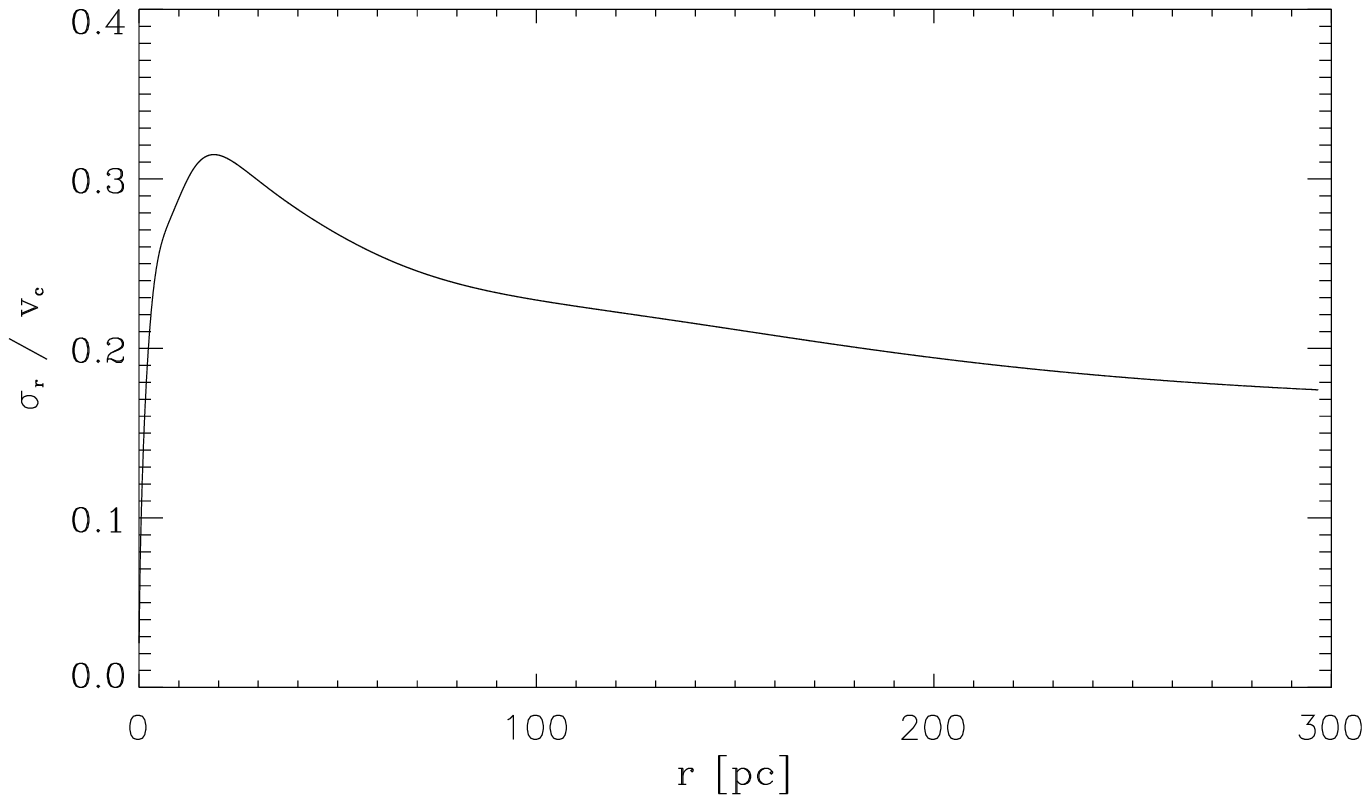,width=8.5cm,clip=}
\caption{The ratio $\sigma_r / v_c$ for the model with
\mbh$=4\cdot10^7$ \msun.}
\label{fig:sigma_ratio}
\end{center}
\end{figure}

We assume that the motions in the gas disc are close to isotropic in
the radial and vertical direction. Then $\sigma_z = \sigma_r$ and
$\left< v_r v_z \right>=0$, and the asymmetric drift correction can be
expressed as
\begin{equation}
v_c^2-v_\phi^2=\sigma_r^2 \left[ -r \frac{d\ln \nu}{dr}-r\frac{d\ln\sigma_r^2}{dr}-\left( 1-\frac{\sigma_r^2}{\sigma_\phi^2} \right)\right]
\end{equation}
where $\sigma_\phi$ is the azimuthal velocity dispersion and $\nu$ is
the number density of gas clouds in the disc (e.g., Binney \& Tremaine
1987). We assume that the number density of the gas clouds is proportional
to the gas flux whose radial profile has been parametrized in Eq.
\ref{eq:flux}. The relation between $\sigma_r$ and $\sigma_\phi$ 
is given by the epicycle approximation
\begin{equation}
\frac{\sigma_\phi^2}{\sigma_r^2}=\frac{-B}{A-B}
\end{equation}
where $A$ and $B$ are the Oort constants given by
\begin{eqnarray}
A &=& \frac{1}{2}\left(\frac{v_c}{r}-\frac{dv_c}{dr} \right)  \\
B &=& -\frac{1}{2}\left(\frac{v_c}{r}+\frac{dv_c}{dr} \right).       
\end{eqnarray}
The approximation is strictly valid in the limit $\sigma \ll v_c$. If
we assume that the SMBH mass is given by the near-infrared 
\mbh$-$\lbulge\ relation,
we have a ratio $\sigma_r/v_c < 0.32$ everywhere in the gaseous disc
of NGC~4435 (Figure \ref{fig:sigma_ratio}) and the method can be
adopted for at least an approximate treatment of the asymmetric drift.
After calculating the asymmetric drift, we obtained the line-of-sight
velocity and velocity dispersion as
\begin{equation}
v = v_\phi \left( \frac{y\sin{i}}{r} \right)
\end{equation}
and 
\begin{equation}
\sigma = (\sigma_\phi^2 - \sigma_r^2) \left( \frac{y\sin{i}}{r} 
\right)^2 + \sigma_r^2,
\end{equation}
respectively. These values of $v$ and $\sigma$ are used in
Eq. \ref{eq:model} for the model calculation. For models without an
asymmetric drift, the above equations reduce to $v_c=v_\phi$ and
$\sigma=\sigma_r=\sigma_\phi$, corresponding to the case where the
intrinsic velocity dispersion represents a locally isotropic
turbulence within the gaseous disc.

\begin{figure} 
\begin{center}
\psfig{file=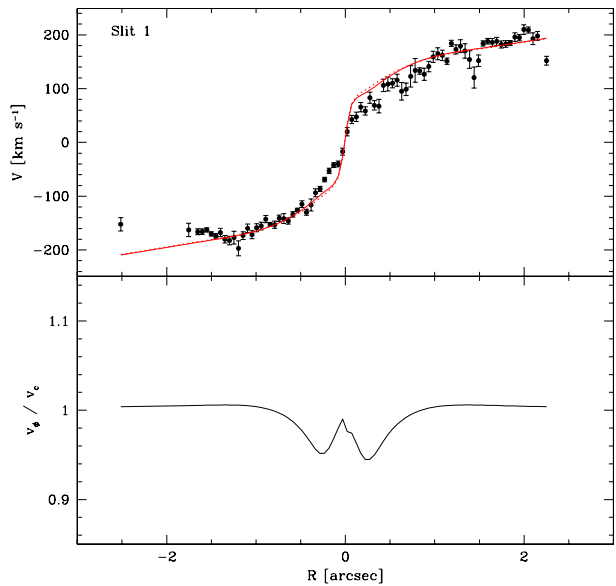,width=8.5cm,clip=}
\caption{{\it Upper panel:\/} The observed \niig\ ({\it filled circles\/}) 
  along the the major axis of NGC~4435 is compared to the one obtained
  with \mbh$ = 4\cdot10^7$ \msun , \mlstar$ = 2.2$ \mlsun , $\theta =
  17^{\circ}$ and $i=70^{\circ}$ with ({\it solid line\/}) and without
  ({\it dashed line\/}) applying the asymmetric drift correction. 
  {\it Lower panel:\/} Ratio between the rotation curves obtained
  with and without applying the asymmetric drift correction.}
\label{fig:drift_correction}
\end{center} 
\end{figure}

We applied the asymmetric drift correction for \mbh$ = 4 \cdot 10^7$
\msun\ and found that the contribution of intrinsic velocity dispersion 
to the rotation curve is negligible ($\leq5\%$). The comparison
between the model rotation curves obtained along the galaxy major axis
with and without the asymmetric drift correction is shown in Figure
\ref{fig:drift_correction}. We can conclude that the contribution of
the velocity dispersion is not enough to allow the presence of
\mbh $ = 4 \cdot 10^7$ \msun\ in the nucleus of NGC~4435.

\section{Conclusions}
\label{sec:conclusions}

We presented long-slit STIS measurements of the ionised-gas kinematics
in the nucleus of the SB0 NGC~4435. NGC~4435 belongs to a sample of 3
galaxies which were selected as STIS targets on the basis of their
CNKD kinematics (Bertola et al. 1998).

For NGC 4435 we found the following:

\medskip
\noindent
{\it (i)\/} It shows a CNKD with a symmetric and regular ionised-gas
kinematics, which is suitable for dynamical modeling in order to
estimate the mass of the central SMBH.  Moreover, NGC~4435 is
characterized by the presence of smooth and circularly symmetric dust
lanes.
This result is in agreement with early findings of Ho et al. (2002)
who found an empirical correlation between the morphology of the dust
lanes and the degree of regularity in the gas velocity field.
 
Bertola et al. (1998) demonstrated that it is possible to detect the
signature of a CNKD in the emission-line PVDs obtained from
ground-based spectroscopy of nearby galaxies. Using this result, Funes
et al. (2002) estimated that the frequency of CNKDs in randomly
selected emission-line disc galaxies is $\la20\%$. However, this
criterion has to be combined with that of the presence of a regular
dust-lane morphology (Ho et al. 2002).
On the basis of our STIS observations, we conclude that less than
$10\%$ of nearby disc galaxies, which show narrow emission lines in
their ground-based spectra, host a CNKD with a velocity field which is
useful for dynamical modeling at \hst\ resolution.

\medskip
\noindent
{\it (ii)\/} We modeled the ionised-gas velocity field of NGC~4435 by
assuming the gas is moving onto circular orbits in an infinitesimally
thin disc located around the SMBH. We constrained the orientation of
the gaseous disc ($\theta = 17^\circ\pm1^\circ$,
$i=70^\circ\pm2^\circ$) by assuming that the morphology of dust lanes
is tracing the gas distribution.
The comparison between the observed and modeled velocity field has
been done by taking into account several effects related to telescope
and instrument optics (nonzero aperture size, apparent wavelength
shifts for light entering the slit off centre, and anamorphic
magnification) as well as detector readout (charge bleeding between
adjacent CCD pixels). We derived for the SMBH mass of NGC~4435 an
upper limit $M_\bullet\leq 7.5\cdot10^6$ \msun\ at 3$\sigma$
confidence level. 

\medskip
\noindent
{\it (iii)\/} The upper limit we derived for the SMBH mass of NGC~4435
considerably falls short of the value of \mbh $ = 4 \cdot 10^7$ \msun\
predicted by the \mbh$-$\sigmac\ relation (Ferrarese \& Ford 2005), as
well as of the value of \mbh $ = 5 \cdot 10^7$ \msun\ predicted by the
near-infrared \mbh$-$\lbulge\ relation (Marconi \& Hunt 2003) as shown
in Figure \ref{fig:scaling_relations}. This discrepancy is not due to
the noise in our spectra (Section 6.1), is robust against
uncertainties in the disc orientation (Section 6.2) or slit
positioning (Section 6.3), and cannot be ascribed to the presence of
nuclear dust (Section 6.4) or pressure-supported gas (Section 6.5).
The measured velocity gradient is consistent with a \mbh$ = 4 \cdot
10^7$ \msun\ only if we assume an ad hoc orientation ($\theta =
10^\circ\pm1^\circ$, $i=28^\circ\pm2^\circ$) for the central portion
of the gaseous disc ($|r| \leq 0.25$ arcsec). However, the images do
not suggest such a dramatic variation of the disc orientation in the
observed radial range.

\medskip
\noindent
{\it (iv)\/} The case of NGC~4435 is very similar to that of the
elliptical NGC~4335 (Verdoes Kleijn et al. 2002), where the ionised-gas
kinematics measured with STIS along three parallel positions is
consistent with that of a CNKD rotating around a SMBH with an
unusually low mass ($M_\bullet\leq 10^8$ \msun ) for its high velocity
dispersion (\sigmac $= 300\pm25$ \kms , see
Figure \ref{fig:scaling_relations}).
Both NGC~4435 and NGC~4335 are amenable to gas dynamical
modeling. Their nuclear discs of ionised gas have a well-constrained
orientation as well as a symmetric and regular velocity field. There
is no evidence that gas is affected by non-gravitational motions or
that it is not in equilibrium, as observed for instance in the E3
galaxy IC 1459 (Cappellari et al. 2002) and the Sbc NGC~4041 (Marconi
et al. 2003).
For NGC~4335 the observed high gas velocity dispersion in the central
arcsec potentially invalidates the \mbh\ measurement based on a thin
rotating disc model leading to a larger \mbh\ value and to a smaller
discrepancy with the \mbh\ scaling laws. On the contrary, in NGC~4435
the more limited width of the lines allowed us to consider the
possibility of a modest dynamical support for the gas clouds.  This
has little impact on the measured \mbh\ and can not explain its
discrepancy with the values predicted by the \mbh\ scaling relations.
As such, if we suppose the resulting upper limit for the SMBH mass of
NGC 4435 is also unreliable, we are then forced to conclude that all
the SMBH masses derived from gaseous kinematics following standard
assumptions have to be treated with caution, even if they agree with
the predictions of the scaling relations.

\smallskip

Thus, NGC~4435 is so far the best candidate of a galaxy with a massive
bulge component with a lower \mbh\ content than predicted by the
\mbh$-$\sigmac\ and \mbh$-$\lbulge\ scaling relations.
Nevertheless, there is an increasing evidence of galaxies for which
dynamical models impose an upper limit to the SMBH mass which is
either lower than or marginally consistent with the one predicted by
the \mbh$-$\sigmac\ (Sarzi et al. 2001; Merritt et al. 2001; Gebhardt
et al. 2001; Valluri et al. 2005). These objects may represent the
population of laggard SMBHs with masses below the \mbh$-$\sigmac\
relation as discussed by Vittorini et al. (2005). Laggard SMBHs are
expected to reside in galaxies that spent most of their lifetime in
the field, where encounters are late and rare so as to cause only slow
gas fueling of the galactic centre and a limited growth of the SMBH
mass.

In light of the serious implications on the slope and intrinsic
scatter of the \mbh\ relations that our results would imply, an
independent measurement of the SMBH mass NGC~4435 based on stellar
dynamical modeling is highly desirable. This could be done with
near-infrared spectroscopic observations with 8m-class telescopes
assisted by adaptive optics systems. Recently, Houghton et al. (2005)
have shown in the case of the giant elliptical NGC~1399 that this
technique can deliver diffraction limited high signal-to-noise spectra
suitable for measuring the stellar LOSVD within the
sphere-of-influence of the SMBH, providing a concrete alternative to
\hst\ that today is in serious jeopardy.

\bigskip
\noindent
{\bf Acknowledgements.}

\noindent

EDB acknowledges the Fondazione ``Ing. Aldo Gini'' for a research
fellowship, and the Herzberg Institute of Astrophysics, Victoria, BC,
for the hospitality while this paper was in progress.
We wish to thank Aaron Barth and Laura Ferrarese for stimulating
discussions, Alfonso Cavaliere and Valerio Vittorini for providing us
their results about laggard SMBHs prior to publication.
We are indebted to Jairo Mendez Abreu for the package which we used
for measuring the photometric parameters of NGC 4435. This research
has made use of the Lyon-Meudon Extragalactic Database (LEDA) and of
the NASA/IPAC Extragalactic Database (NED).
We would like to thank the anonymous referee for constructive
comments.

\appendix 

\section{Calculation of the color excess $E(B-I)$ in the case of a \mbh$ = 4 \cdot 10^7$ \msun}

In this section we evaluate the colour excess $E'(B-I)$ we would
expect in the case that a \mbh$ = 4 \cdot 10^7$ \msun\ is present and
the absorption of the central dust is able to reproduce the observed
kinematics (see Section 6.4).

\begin{figure}
\psfig{file=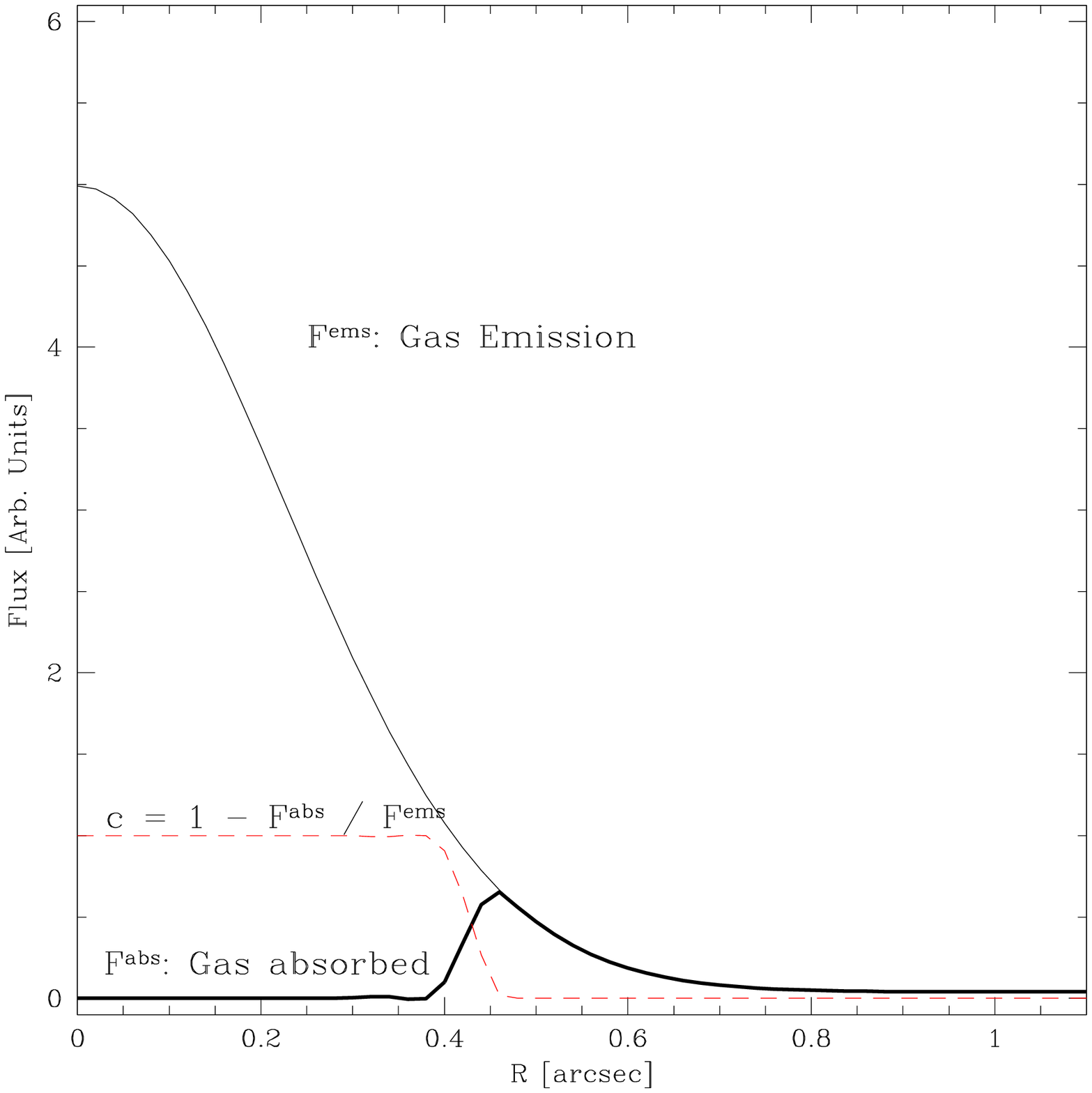,width=8.5cm,clip=}
\caption{Intrinsic flux radial profile of the gas in the NGC 4435
  model. The {\it solid thin line} shows the intrinsic emission of our
  best model, whereas the {\it solid thick line} shows the absorbed
  emission needed to accomodate a \mbh$=4\cdot10^7$ \msun\ and match
  the observed velocities (Figure 17). The {\it dashed line} shows the
  efficiency of the dust absorption, $c$, as defined in the text.}
\label{fig:app1}
\end{figure}

If the ionised gas is settled in the equatorial plane as well as the
dust, the intrinsic flux corrected by absorption ($F^{abs}$) will be
related to the intrinsic one ($F^{ems}$) by the relation
\begin{equation}
F^{abs} = 10^{-0.4 A_{[NII]}} \cdot F^{ems} = (1-c)\cdot F^{ems}
\end{equation}
The coefficient $c$ indicates the efficiency of the absorption, and it
can be easily derived (see Figure \ref{fig:app1}). For $c=1$,
$F^{abs}=0$, for $c=0$, $F^{abs}=F^{ems}$.

Using the standard Galactic extinction (Cardelli et al. 1989) we can
determine the relation between the extinction $A_{[NII]}$ and the
extinction in the I band, $A_I$:
\begin{equation}
A_I = 0.588 A_{[NII]}
\end{equation}
The central stellar component (bulge plus disc) is thicker with
respect the ionised gas and the dust discs. Thus, only the portion of
stars behind the dust layer is absorbed:
\begin{equation}
 F_{star}^{abs} = \frac{1}{2}F_{star}^{ems} +\frac{1}{2}10^{-0.4\cdot A_I}F_{star}^{ems} =\frac{1+10^{-0.4\cdot A_I}}{2}F_{star}^{ems} 
\end{equation}
What we should observe is the extinction $A'_I$ defined by:
\begin{equation}
F_{star}^{abs} = 10^{-0.4\cdot A'_I} \cdot F_{star}^{ems} 
\end{equation}
with (comparing Eqs. A3 and A4):
\begin{equation}
A'_I = -2.5 \log \frac{1+10^{-0.4\cdot A_I}}{2}
\end{equation}
Then, according to the standard Galactic extinction (Cardelli et
al. 1989), the expected color excess $E'(B-I)$ is:
\begin{equation}
E'(B-I)= A'_I/0.5583
\end{equation}
Then, combining Eqs. A1, A2, A5 and A6, it is possible to calculate the map of $E'(B-I)$ as a function of
$c$, derived from Figure \ref{fig:app1}.
\begin{equation}
E'(B-I)= -4.48 \log \frac{1+(1-c)^{0.588}}{2}
\end{equation}
In the central region, where $c=1$ the expected color excess is 1.3
mag, while the observed one is 0.6 mag. (Figure 5).

\bsp

\label{lastpage} 

\bigskip
\noindent

\begin{thebibliography}{}


\bibitem[xxx]{bae} Baes, M., Buyle, P., Hau, G. K. T., \&\ Dejonghe,
H., 2003, MNRAS, 341, L44

\bibitem[xxx]{bar00} Barth, A. J. 2004, in Coevolution of Black Holes \&\ Galaxies, ed. L. C. Ho (Cambridge: Cambridge Univ. Press), 21

\bibitem[xxx]{bar} Barth, A. J., Sarzi, M., Rix, H,-W, Ho, L. C., Filippenko, A. V.,\&
Sargent, W. L. W. 2001, ApJ, 555, 685

	
\bibitem[xxx]{ber} Bernardi, M., Alonso, M. V., da Costa, L. N., et al. 2002, AJ, 123, 2990

\bibitem[xxx]{ber95}Bertola, F.,  Cinzano, P.,  Corsini, E. M.,  Rix, H.-W., \& Zeilinger, W. W. 1995, ApJ, 448, L13

\bibitem[xxx]{ber00} Bertola, F.,
 Cappellari, M., Funes, J.~G., Corsini, E.~M., Pizzella, A., \&\ Vega
 Beltr\'an, J.~C.\ 1998, ApJ, 509, L93

\bibitem[xxx]{bin}Binney, J., \& Tremaine, S. 1994, Galactic Dynamics (Princeton University Press: Princeton)

\bibitem[xxx]{bow94} Bowers, C., \&\ Baum, S. 1998, STIS Instrument Science Rep. 98-23 (Baltimore: STScI)

\bibitem[xxx]{bro} Brown, T. et al. 2002, in HST STIS Data Handbook, version 4.0, ed. B. Mobasher, Baltimore, STScI

\bibitem[xxx]{cap} Cappellari, M., et al. 2002, ApJ, 578, 787

\bibitem[xxx]{car}Cardelli, J. A., Clayton, G. C., \& Mathis, J. S.
1989, ApJ, 345, 245

\bibitem[xxx]{cin} Cinzano, P., Rix, H.-W., Sarzi, M., Corsini, E. M., Zeilinger, W. W., \& Bertola, F. 1999, MNRAS, 307, 433


\bibitem[xxx]{dev} de Vaucouleurs, G., de
 Vaucouleurs, A., Corwin, H.\ G.\ Jr., Buta, R.\ J., Paturel, G., \&
 Fouqu\`e, P.  1991, Third Reference Catalogue of Bright Galaxies
 (New York: Springer-Verlag) (RC3)

\bibitem[xxx]{fer01}Ferrarese, L.,  2002, ApJ, 578, 90

\bibitem[xxx]{fer02} Ferrarese, L., \&\ Merritt, D.,  2000, ApJ, 539, L9


\bibitem[xxx]{fer03} Ferrarese, L., \& Ford, H. 2005, Space Science Reviews, 116, 523

\bibitem[xxx]{fill} Fillmore, J. A., Boroson, T. A., \& Dressler, A. 1986, ApJ, 302, 208

\bibitem[xxx]{fun} Funes, J. G., S. J, Corsini, E. M.,
 Cappellari, M., Pizzella, A., Vega Beltr\'an, J. C., Scarlata, C.,
 \& Bertola, F. 2002, A\&A, 388, 50

\bibitem[xxx]{geb00} Gebhardt K., et al., 2000, ApJ, 539, 13

\bibitem[xxx]{geb01} Gebhardt K., et al., 2001, AJ, 122, 2469 

\bibitem[xxx]{ghe01} Ghez, A.M., et al. 2003, ApJ, 586, L127

\bibitem[xxx]{gra00} Graham, J. A., Ferrarese, L., Freedman, W. L., et
 al. 1999, ApJ, 516, 626

\bibitem[xxx]{gra01} Graham, A. W., Erwin, P., Caon, N., \& Trujillo, I. 2001, ApJ, 563, L11

\bibitem[xxx]{gut} Guthrie, B.\ N.\ G. 1992, A\&AS, 93, 255

\bibitem[xxx]{har} H\"aring, N., \& Rix, H.-W. 2004, ApJ, 604, L89

\bibitem[xxx]{ho} Ho, L. C., et al. 2002, PASP, 114, 137

\bibitem[xxx]{hol} Holtzman, J. A., Burrows, C. J., Casertano, S.,
Hester, J. J., Trauger, J. T., Watson, A. M., \& Worthey, G.  1995,
PASP, 107, 1065

\bibitem[xxx]{hou} Houghton, R. C. W., Magorrian, J., Sarzi, et
al. 2005, MNRAS in press (astro-ph/0510278)

\bibitem[xxx]{jar} Jarrett, T. H., Chester, T., Cutri, R., Schneider,
S., Skrutskie, M., \& Huchra, J. P. 2000, AJ, 119, 2498

\bibitem[xxx]{jor} Jorgensen, I., Franx, M., \&\ Kjaergaard, P., 1995,
MNRAS, 276, 1341

\bibitem[xxx]{kin} Kinney, A. L., Calzetti, D., Bohlin, R. C., McQuade, K., Storchi-Bergmann, T., \& Schmitt, H. R.  1996, ApJ, 467, 38

\bibitem[xxx]{kor00} Kormendy, 2001, Rev. Mexicana Astron. Astrofis,
Ser. Conf., 10, 69

\bibitem[xxx]{kor01} Kormendy, J. 2004, in Coevolution of Black Holes
\& Galaxies, ed. L. C. Ho (Cambridge: Cambridge Univ. Press), 1

\bibitem[xxx]{kor} Kormendy, J. \&  Richstone, D. 1995, ARA\&A, 33, 581

\bibitem[xxx]{kry} Krist, J., \& Hook, R. 1999, The Tiny Tim User's Guide (Baltimore: STScI)

\bibitem[xxx]{mac} Maciejewski, W.,\& Binney, J. 2001, MNRAS, 323, 831

\bibitem[xxx]{mag} Magorrian, J., et al.  1998, AJ, 115, 2285

\bibitem[xxx]{mar} Marconi, A., \& Hunt, L. K. 2003, ApJ, 589, 21

\bibitem[xxx]{mar01} Marconi, A., et al. 2003, ApJ, 586, 868

\bibitem[xxx]{med04} Mendez Abreu, J., Corsini, E. M., Aguerri, J. A. L.
 2004, in Baryons in Dark Matter Halos, ed. R. Dettmar, U. Klein, \& P. Salucci, PoS (Trieste: SISSA), p.83.1

\bibitem[xxx]{mer} Merritt D., Ferrarese L., Joseph C.~L., 2001, Sci, 293, 1116 
 
\bibitem[xxx]{miy} Miyoshi, M., Moran, J., Herrnstein, J., Greenhill, L., Nakai, N., Diamond, P., \& Inoue, M. 1995, Nature, 373, 127

\bibitem[xxx]{mon} Monnet, G., Bacon, R., \& Emsellem, E.  1992, A\&A, 253, 366


\bibitem[xxx]{pig} Pignatelli, E., Corsini, E. M., Vega B\'eltran, J. C., Scarlata, C., et al. 2001, MNRAS, 320, 124

\bibitem[xxx]{piz0} Pizzella, A., Corsini, E.~M., Morelli, L., Sarzi,
M., Scarlata, C., Stiavelli, M., \& Bertola, F.\ 2002, ApJ, 573, 131

\bibitem[xxx]{piz} Pizzella, A., Corsini, E. M., Dalla Bont\`a, E., Sarzi M., Coccato, L., \& Bertola, F. 2005, ApJ, 631, 785

\bibitem[xxx]{pog} Pogge, R. W., Maoz, D., Ho, L. C., \& Eracleous, M. 2000, ApJ, 532, 323

\bibitem[xxx]{pre} Press, W. H., Teukolsky, S. A., Vetterling, W. T.,
\&\ Flannery, B. P. 1992, Numerical Recipes in Fortran 77 (2d ed.:
Cambridge: Cambridge University Press)

\bibitem[xxx]{rub} Rubin, V. C., Kenney, J. D. P., \& Young, J. S.  1997, AJ, 113, 1250

\bibitem[xxx]{sar} Sarzi, M., et al. 2001, ApJ, 550, 65

\bibitem[xxx]{sim} Simien, F., \& Prugniel, Ph. 1997, A\&AS, 126, 15

\bibitem[xxx]{sof} Sofue, Y., Tomita, A., Tutui, Y. Honma, M.,\& Takeda, Y. 1998, PASJ, 50, 427

\bibitem[xxx]{ton} Tonry, J.\& Davis, M. 1981, ApJ, 246, 666

\bibitem[xxx]{val} Valluri, M., Ferrarese, L., Merritt D., Joseph
C.~L., 2005, ApJ, 628, 137

\bibitem[xxx]{van} van Dokkum, P. G. 2001, PASP, 113, 1420

\bibitem[xxx]{ver} Verdoes Kleijn, G. A., van der Marel, R. P., de
Zeeuw, P. T., Noel-Storr, J., Baum, S. A. 2002, AJ, 124, 2524

\bibitem[xxx]{vit}Vittorini, V., Shankar, F.,\& Cavaliere, A. 2005, MNRAS, in press (astro-ph/0508640)

\bibitem[xxx]{whi} Whitmore, B. 1995, in Calibrating Hubble Space Telescope: Post Servicing Mission, ed. A. Koratkar \& C. Leitherer (Baltimore: STScI), 269

\end{thebibliography}
\end{document}